\documentclass[amssymb,twocolumn,aps]{revtex4-2}
\usepackage{graphicx}
\usepackage{enumerate}
\usepackage{amssymb}
\usepackage{amsmath}
\usepackage{amsfonts}
\usepackage{color}
\usepackage{mathrsfs}

\begin{document}

\title{Fast electrons interacting with chiral matter: mirror symmetry breaking of quantum decoherence and lateral momentum transfer}

\author{A. Ciattoni$^1$}
\email{alessandro.ciattoni@spin.cnr.it}
\affiliation{$^1$CNR-SPIN, c/o Dip.to di Scienze Fisiche e Chimiche, Via Vetoio, 67100 Coppito (L'Aquila), Italy}

\date{\today}

\begin{abstract}
Photons experience mirror asymmetry of macroscopic chiral media, as in circular dichroism and polarization rotation, since left and right handed circular polarizations differently couple with matter handedness. Conversely, free relativistic electrons with vanishing orbital angular momentum have no handedness so the question arises whether they could sense chirality of geometrically symmetric macroscopic samples. In this Letter, we show that matter chirality breaks mirror symmetry of the scattered electrons quantum decoherence, even when the incident electron wave function and the  sample shape have a common reflection symmetry plane. This is physically possible since the wave function transverse smearing triggers electron sensitivity to the spatial asymmetry of the electromagnetic interaction with the sample, as results from our non-perturbative analysis of the scattered electron reduced density matrix, in the framework of macroscopic quantum electrodynamics. Furthermore, we prove that mirror asymmetry also shows up in the distribution of the electron lateral momentum, orthogonal to the geometric symmetry plane, whose non-vanishing mean value reveals that the electron experiences a lateral mechanical interaction entirely produced by matter chirality.
\end{abstract}

\maketitle
Molecular chirality is an important topic in science both for its implications in chemistry, biology, and medicine and for the physical effects produced by microscopic mirror symmetry breaking. The coupling of a chiral molecule with the electromagnetic field involves both its electric and magnetic dipoles \cite{Berov} whose combined polar and axial characters trigger molecular sensitivity to radiation handedness, a property routinely used to detect chirality by means of circular dichroism and polarization rotation \cite{Barro}. The large mismatch between molecular size and wavelength both makes such effects very small and establishes the dipolar interaction which forbids the molecule to sense the spatial variations of the field. However, macroscopic chiral samples with a large number of chiral molecules display macroscopic mirror symmetry breaking and consequently they support chiroptical effects driven by the symmetry of the spatial field profile, as the suppression of field mirror symmentry upon reflection by chiral molecular films (mirror optical activity) \cite{Ciat1,Ciat2}, or the antenna emission of asymmetric fields in chiral metamaterials \cite{Hisam}. Tight subwavelength confinement of the field reduces the molecule-field spatial scale mismatch and accordingly it enhances the above chiroptical spatial effects, in analogy to other chiral nonanophotonic phenomena \cite{Muuuu,Govor,Liuuu,Abdul,Neste}. 

The ability of electron microscopes to focus the beam down to the subnanometer scale makes fast electrons an ideal continuum source of ultra-confined electromagnetic field \cite{deAb1} whose evanescent character can in principle be used to enhance spatially asymmetric chiroptical effects, this leading to indirect chiral sensing by detecting the diffraction radiation from a lateral nanostructure \cite{Ciat1}. On the other hand, only few methods have been proposed to directly sense chirality in electron microscopes. Magnetically induced chirality can be probed by comparing electron energy loss spectra taken in opposite magnetic fields and under suitable electron scattering conditions (electron energy-loss magnetic chiral dichroism) \cite{Scha1,Scha2}. In the nonmagnetic case, the idea of a probing electron with an handedness able to sense matter chirality, has led to consider electron vortex beams \cite{Verbe,Bliok} carrying orbital angular momentum whose exchange with the sample has been shown to provide chiral information in both elastic scattering from crystals \cite{Juch1,Juch2} and inelastic scattering from plasmonic samples with chiral shape and single chiral molecules \cite{Asenj}. A different approach based on the photon-induced near-field electron microscopy technique \cite{Barwi,Parkk,Vanac} has been proposed \cite{Harve} where the probing handedness is injected by means of circularly polarized radiation and the chirality of the plasmonic sample shape is retrieved from the spatial scan of the electron energy spectrum.

\begin{figure}
\centering
\includegraphics[width = 1\linewidth]{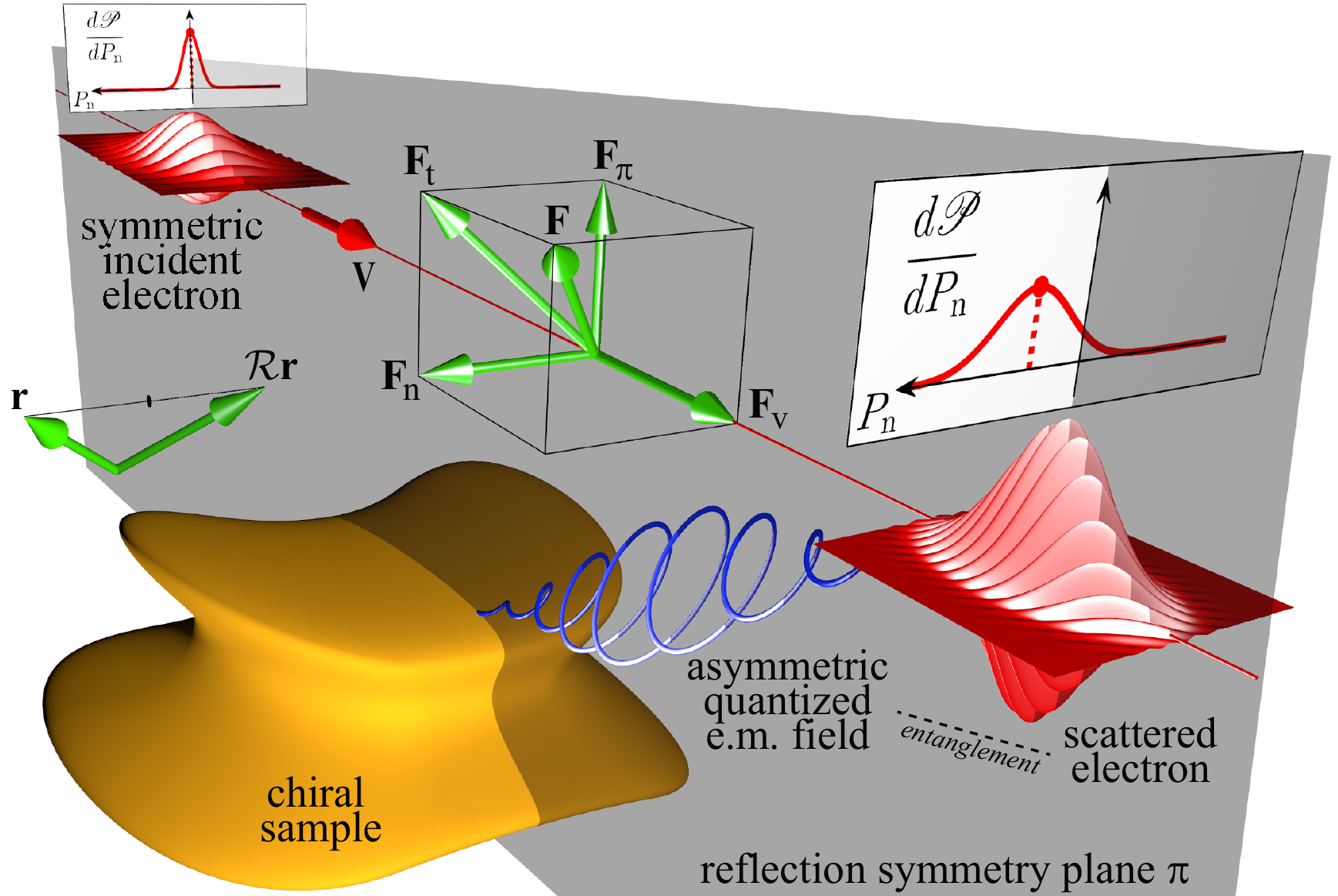}
\caption{Scattering setup. The shape of the chiral sample has a reflection symmetry plane $\pi$ ($\cal R$ is the reflection through $\pi$) and the incident electron wave function is mirror symmetric with its velocity $\bf V$ lying on $\pi$. The scattered electron is entangled with the spatially asymmetric quantized electromagnetic field mediating its interaction with the sample, this producing mirror symmetry breaking of electron decoherence. ${\bf{F}}_{\rm v}$ and ${\bf{F}}_{\rm t}$ are the parts of a vector $\bf F$ parallel and orthogonal to $\bf V$ whereas ${\bf{F}}_{\rm \pi}$ and ${\bf{F}}_{\rm n}$ are the parts of ${\bf{F}}_{\rm t}$ parallel and orthogonal to $\pi$. The distribution $d \mathscr{P}/d P_{\rm n}$ of the lateral momentum $P_{\rm n}$ of the scattered electron is not symmetric as opposed to the incident electron one.}
\label{Figure1}
\end{figure}

In this Letter, we show that a fast electron with initial mirror symmetric wave function, i.e. with no handedness, does experience microscopic chirality of a symmetrically shaped macroscopic sample (see Fig.1): the quantum decoherence of the scattered electron turns out to be mirror asymmetric due to a chiroptical spatial effect generalizing mirror optical activity. To this end, we evaluate the non-perturbative reduced density matrix of the scattered electron by using macroscopic quantum electrodynamics \cite{Grune,Schee,Yoshi,Butch}, an appropriate framework to deal with the electron-sample entanglement \cite{Hayun,Karni,Kfirr} which is ultimately responsible for electron decoherence \cite{Fordd,Hsian,Scha3,Meche}. Besides, by using the density matrix, we show that the distribution of the momentum component orthogonal to the reflection symmetry plane (see Fig.1) is not mirror symmetric thus showing that the scattered electron gains a net lateral momentum purely due to the sample chirality. We specialize our general discussion to an aloof electron probing a realistic chiral nanofilm where the above asymmetric effects are magnified by the relatively long effective interaction length.

Macroscopic quantum electrodynamics provides the description of the quantized field coupled to the chiral sample via the spectral electric field ${\bf{\hat E}}_\omega   ( {\bf{r}}  ) = \sum\limits_\lambda  {\int {d^3 {\bf{r}}'} {\cal G}_{\omega \lambda }  ( {{\bf{r}},{\bf{r}}'}  ){\bf{\hat f}}_{\omega \lambda }  ( {{\bf{r}}'}  )}$ and Hamiltonian $\hat H_{em}  = \int\limits_0^\infty  {d\omega } \hbar \omega \sum\limits_\lambda  {\int {d^3 {\bf{r}}} \, {\bf{\hat f}}_{\omega \lambda }^\dag   ( {\bf{r}}  ) \cdot {\bf{\hat f}}_{\omega \lambda }  ( {\bf{r}}  )}$  where $\hat {\bf f}_{\omega \lambda}   ( {\bf{r}}  ) \cdot {\bf e}_j$ is the annihilation operator of a polaritonic excitation of frequency $\omega$ kind $\lambda = e,m$ (electric and magnetic) position $\bf r$ and direction $j$, whereas ${\cal G}_{\omega \lambda} ( {{\bf{r}},{\bf{r}}'} )$ are two tensors related to the sample Green's tensor ${\cal G}_\omega ( {{\bf{r}},{\bf{r}}'} )$ \cite{Yoshi} (see Sec.S1 of Supplemental Material) in turn satisfying the inhomogeneous Helmholtz equation
\begin{eqnarray} \label{Green}
 \left[ \nabla  \times \frac{1}{\mu }\nabla  \times  - \frac{{\omega ^2 }}{{c^2 }}\left( {\varepsilon  - \frac{{\kappa ^2 }}{\mu }} \right) + \frac{\omega }{c}\left( {\nabla \frac{\kappa }{\mu }} \right) \times  \right. + \nonumber \\ 
 \left.  + 2 \frac{\omega }{c}\frac{{\kappa }}{\mu }\nabla  \times  \right]{\cal G}_\omega   ( {{\bf{r}},{\bf{r}}'}  ) = \delta  ( {{\bf{r}} - {\bf{r}}'}  )I 
\end{eqnarray}
where $\varepsilon({\bf r},\omega)$, $\mu({\bf r},\omega)$ and $\kappa({\bf r},\omega)$ are the medium permittivity, permeability and chiral parameter, respectively. The electron  is assumed to be almost monoenergetic and highly directional with its momentum $\bf P$ slightly differing from a reference momentum ${\bf P}_0$ so that its paraxial Hamiltonian with no momentum recoil is $\hat H_{el}  = {\bf{V}} \cdot {\bf{\hat P}}$ where ${\bf{V}} = c{\bf{P}}_0 /\sqrt {m^2 c^2  + P_0^2 }$ and $\bf {\hat P}$ are the electron velocity and momentum operator. In the same approximation, the chiral sample-electron interaction hamiltonian is ${\hat H_{int} } ={e [ {{\bf{V}} \cdot {\bf{\hat A}} ( {{\bf{\hat R}}}  ) - \hat \Phi  ( {{\bf{\hat R}}}  )}  ]}$ where $-e<0$ and $\bf{\hat R}$ are the electron charge and position operator whereas ${{\bf{\hat A}}}$ and ${\hat \Phi }$ are the vector and scalar potential operators in the Coulomb gauge related to the transverse and longitudinal parts of the electric field operator. Time evolution produced by the Hamiltonian $\hat H_{em} + \hat H_{el} + \hat H_{int}$ can be analitically dealt with, in analogy to the achiral situation \cite{Rive1}, and the rigorous scattering operator $\hat S$ we have obtained is  detailed in Sec.S2.2 of Supplemental Material. The incident electron wave function is $\psi _i  ( {\bf{R}}  ) = \phi _i  ( {{\bf{R}}_{\rm t} }  )e^{i\frac{{E_i }}{{\hbar V}}R_{\rm v} } / {\sqrt \ell  }$ where subscripts $\rm v$ and $\rm t$ henceforth label the parts of a vector parallel and orthogonal to $\bf V$ (see Fig.1), $\phi _i$ is the initial transverse wave function profile, $E_i$ is the initial electron energy and $\ell$ is the longitudinal quantization length. Since no radiation is initially involved in the scattering setup we are considering, the initial electron-field state is $\left| {\Psi _i } \right\rangle  = \left| {\psi _i } \right\rangle  \otimes \left| 0 \right\rangle$  where $\left| 0 \right\rangle$ is the polaritonic vacuum state. The final state $\left| \Psi  \right\rangle  = \hat S\left| {\Psi _i } \right\rangle$ displays electron-field entanglement, it resulting from processes of electron energy loss and multi-polariton excitation of any order, so that measurement predictions about the electron (with the field left unmeasured) have to be performed by resorting to the reduced density operator $\hat \rho _e  = {\rm Tr}_{\rm Fock} \left| \Psi  \right\rangle \left\langle \Psi  \right|$ where the trace is taken over the polariton Fock space. In the position representation, we get the reduced density matrix (RDM) $\rho _e  ( {{\bf{R}},{\bf{R}}'}  ) = \psi _i  ( {\bf{R}}  )\psi _i^*  ( {{\bf{R}}'}  )\gamma  ( {{\bf{R}},{\bf{R}}'}  )$ (see Sec.S2.3 of Supplemental Material) where 
\begin{equation} \label{gam}
\gamma  ( {{\bf{R}},{\bf{R}}'}  ) = e^{i [ {\Phi  ( {{\bf{R}}_{\rm t} }  ) - \Phi  ( {{\bf{R}}'_{\rm t} }  )}  ] - \frac{1}{2} [ {\Delta  ( {{\bf{R}}_{\rm t} ,{\bf{R}}_{\rm t} }  ) + \Delta  ( {{\bf{R}}'_{\rm t} ,{\bf{R}}'_{\rm t} }  )}  ]} e^{\Delta  ( {{\bf{R}},{\bf{R}}' }  )} 
\end{equation}
is the decoherence factor (DF) of the scattered electron with
\begin{eqnarray} \label{PhiDelta}
 \Phi ( {{\bf{R}}_{\rm t} } ) &=& \frac{{4\alpha }}{c}  \int\limits_0^{ + \infty } {d\omega } \int\limits_{ - \infty }^{ + \infty } {dq} \int\limits_{ - \infty }^q {dq'\,} \sin \left[\frac{\omega }{V}( {q - q'} )\right]  \cdot \nonumber\\
 &\cdot& {\mathop{\rm Im}\nolimits} \left[ {\bf{u}}_{\rm v}  \cdot {{\cal G}_\omega  ( {{\bf{R}}_{\rm t}  + q{\bf{u}}_{\rm v} ,{\bf{R}}_{\rm t}  + q'{\bf{u}}_{\rm v} } )} {\bf{u}}_{\rm v} \right], \nonumber \\
\Delta  ( {{\bf{R}},{\bf{R}}'}  ) &=& \frac{{4\alpha }}{c}\int\limits_0^\infty  {d\omega } e^{ - i\frac{\omega }{V}( {R_{\rm v}  - R'_{\rm v} } )} \int\limits_{ - \infty }^{ + \infty } {dq} \int\limits_{ - \infty }^{ + \infty } {dq'} \,e^{i\frac{\omega }{V} ( {q - q'} )} \cdot \nonumber\\
 &\cdot&  {\mathop{\rm Im}\nolimits} \left[ {\bf{u}}_{\rm v}  \cdot {{\cal G}_\omega   ( {{\bf{R}}_{\rm t}  + q{\bf{u}}_{\rm v} ,{\bf{R}}'_{\rm t}  + q'{\bf{u}}_{\rm v} }  )} {\bf{u}}_{\rm v}  \right], 
\end{eqnarray}
where $\alpha$ is the fine-structure constant and ${\bf{u}}_{\rm v}  = {\bf{V}}/V$ is the velocity unit vector. The first exponential in the right hand side of Eq.(\ref{gam}) physically arises from elastic processes due to the zero-point fluctuations of the field  whereas the second exponential containing $\Delta ({\bf{R}},{\bf{R}}')$ provides the inelastic contribution to electron decoherence, usual mechanisms in achiral environments \cite{DiGiu,deAb2} that we have here generalized in the presence of chiral matter. From the above RDM it is possible to evaluate the energy distribution and, most importantly for our purposes, the transverse momentum distribution (TMD) (see Sec.S2.4 of Supplemental Material) which is 
\begin{eqnarray} \label{dPdPt}
\frac{{d \mathscr{P}}}{{d^2 {\bf{P}}_{\rm t} }} &=& \frac{1}{{ ( {2\pi \hbar }  )^2 }}\int {d^2 {\bf{R}}_{\rm t} } \int {d^2 {\bf{R}}'_{\rm t} } e^{ - \frac{i}{\hbar }{\bf{P}}_{\rm t}  \cdot  ( {{\bf{R}}_{\rm t}  - {\bf{R}}'_{\rm t} }  )} \cdot \nonumber\\
 &\cdot& \phi _i  ( {{\bf{R}}_{\rm t} }  )\phi _i^*  ( {{\bf{R}}'_{\rm t} }  )\gamma  ( {{\bf{R}}_{\rm t} ,{\bf{R}}'_{\rm t} }  ).
\end{eqnarray}

\begin{figure}
\centering
\includegraphics[width = 1\linewidth]{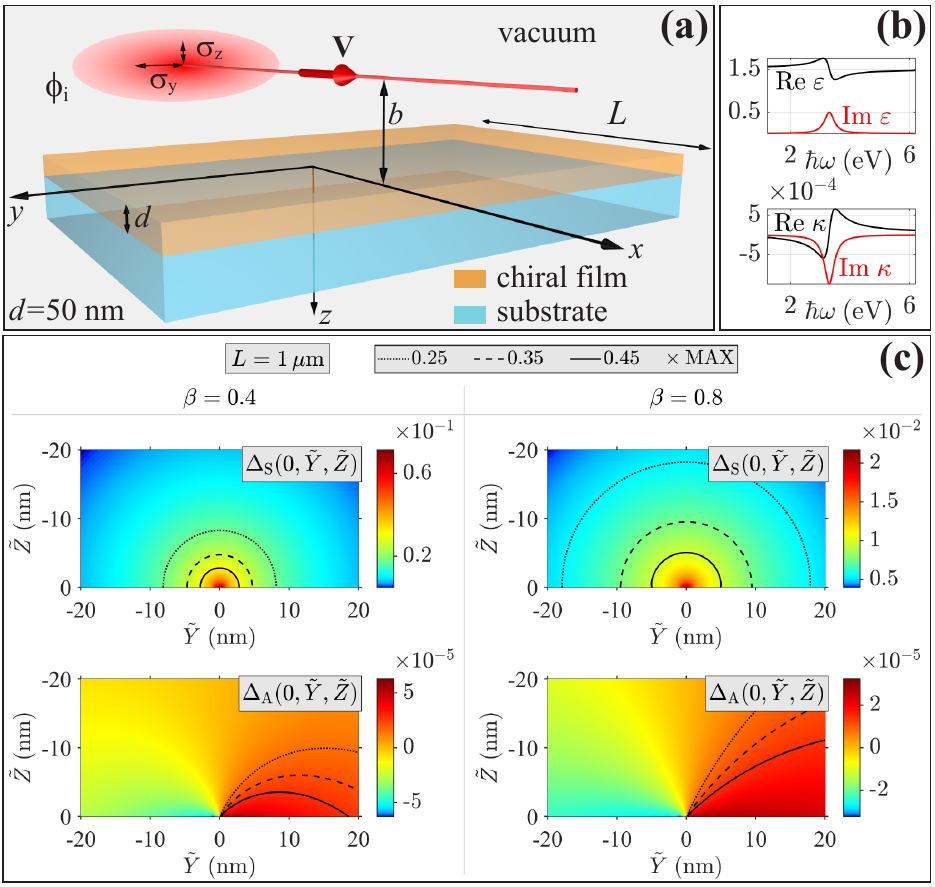}
\caption{(a) Scattering of an aloof electron by a chiral nanofilm deposited on a substrate. (b) Film permittivity $\varepsilon$ and chiral parameter $\kappa$. (c) Symmetric and antisymmetric parts of $\Delta$ for two electron velocities $\beta = V/c$ (columns). For graphical clarity purpose, suitable level curves are plotted as dotted, dashed and solid black lines.}
\label{Figure2}
\end{figure} 

As a prelude to the discussion of the asymmetric features gained by the electron upon scattering, it is essential considering the Green's tensor behavior under spatial reflections. To avoid macroscopic geometrical dissymmetry, possibly hiding microscopic chirality, we consider a chiral sample displaying a mirror symmetry plane $\pi$  and the mirror reflection through it, ${\bf{r}}^{\rm M}  = {\cal R}{\bf{r}}$, so that $ \varepsilon  ( {{\cal R}{\bf{r}}}  ) = \varepsilon \left( {\bf{r}} \right)$, $ \mu  ( {{\cal R}{\bf{r}}}  ) = \mu  ( {\bf{r}}  )$ and  $ \kappa  ( {{\cal R}{\bf{r}}}  ) = \kappa  ( {\bf{r}}  )$ (see Fig.1). Now the mirror image of the Green's tensor is ${\cal G}_\omega ^{\rm M}  ( {{\bf{r}},{\bf{r}}'}  ) = {\cal R}{\cal G}_\omega   ( {{\cal R}{\bf{r}},{\cal R}{\bf{r}}'}  ){\cal R}$ which is easily seen to satisfy Eq.(\ref{Green}) with the sign flip $\kappa \rightarrow - \kappa$, so that ${\cal G}_\omega ^{\rm M}  ( {{\bf{r}},{\bf{r}}'}  )$ is the Green's tensor of the opposite enantiomeric sample, a restatement of reflection invariace of electrodynamics. As a remarkable consequence, ${\cal G}_\omega ^{\rm M}  ( {{\bf{r}},{\bf{r}}'}  )$ can not coincide with ${\cal G}_\omega  ( {{\bf{r}},{\bf{r}}'}  )$ (see Sec.S2.5 of Supplemental Material), since otherwise identical sources in two opposite enantiomeric samples would radiate identical fields, and this yields Green's tensor mirror asymmetry  ${\cal R}{\cal G}_\omega  \left( {{\cal R}{\bf{r}},{\cal R}{\bf{r}}'} \right){\cal R} \ne {\cal G}_\omega  \left( {{\bf{r}},{\bf{r}}'} \right)$. This asymmetry is particularly intriguing in view of its impact on electron dynamics since, from Eq.(\ref{gam}), the transverse features of the Green's tensor are directly conveyed to the DF owing to its local dependence on $({\bf R}_{\rm t},{\bf R}'_{\rm t})$ through ${\cal G}_\omega$. Therefore we consider the maximally symmetric setup of Fig.1 where the electron velocity $\bf V$ lies on $\pi$ and its wave function transverse profile is mirror symmetric, i.e. $\phi _i \left( {{\cal R}{\bf{R}}_{\rm t} } \right) = \phi _i \left( {{\bf{R}}_{\rm t} } \right)$. In this configuration DF mirror asymmetry  
\begin{equation} \label{DFMA}
\gamma \left( {{\cal R}{\bf{R}},{\cal R}{\bf{R}}'} \right) \ne \gamma \left( {{\bf{R}},{\bf{R}}'} \right),
\end{equation}
is a direct consequence of the Green's tensor asymmetry and it is one of the main result of this Letter. The RDM of the scattered electron is therefore mirror asymmetric as well, except for a fully localized initital electron, $\phi _i \left( {{\bf{R}}_{\rm t} } \right) = \delta \left( {{\bf{R}}_{\rm t} } \right)$, since in this case the relevant DF is $\gamma \left( {{\bf{R}}_{\rm v} ,{\bf{R}}'_{\rm v} } \right)$ which is evidently mirror symmetric. RDM mirror symmetry breaking can be physically grasped by noting that the chiral sample, when exposed to the symmetric field produced by the bare electron (singular part of ${\cal G}_\omega$), produces a mirror asymmetric field (reflected part of ${\cal G}_\omega$) through a mechanism generalizing mirror optical activity of chiral films \cite{Ciat1}. Such reaction field self-consistently acts back on the electron and its spatial asymmetry is conveyed to the RDM if the electron wave function has a transverse smearing, since a point-like particle lying on the symmetry plane $\pi$ can not sense the spatial symmetry of the field.

Equation (\ref{DFMA}) directly implies that the TMD in Eq.(\ref{dPdPt}) is not left invariant by the transverse momentum reflection ${\bf{P}}_{\rm t} \to {\cal R}{\bf{P}}_{\rm t}$. As a consequence, using the decomposition ${\bf{P}}_{\rm t}  = {\bf{P}}_\pi + {\bf{P}}_{\rm n}$, were subscripts $\rm \pi$ and $\rm n$ label the parts of a transverse vector parallel and orthogonal to $\pi$ (see Fig.1), we get to the result that the lateral momentum distribution of the scattered electron, $d\mathscr{P}/dP_{\rm n}  = \int {dP_\pi  } d\mathscr{P}/d^2 {\bf{P}}_{\rm t}$ (since $d^2 {\bf{P}}_{\rm t}  = dP_{\rm n} dP_\pi$), is not symmetric in $P_{\rm n}$, as opposed to the incident electron one (see Fig.1) which is symmetric by reflection invariance of $\phi_i$. Accordingly, the scattered electron has a non-vanishing lateral momentum mean value which is given by (see Sec.S2.5 of Supplemental Material)
\begin{equation} \label{Pn}
\left\langle {{\bf{P}}_{\rm n} } \right\rangle  = \int {d^2 {\bf{R}}_{\rm t} } \left| {\phi _i } \right|^2 \left[ {\frac{\hbar }{i}\nabla _{{\bf{R}}_{\rm t} } \log \sqrt {\frac{{\gamma \left( {{\bf{R}}_{\rm t} ,{\bf{R}}'_{\rm t} } \right)}}{{\gamma \left( {{\cal R}{\bf{R}}_{\rm t} ,{\cal R}{\bf{R}}'_{\rm t} } \right)}}} } \right]_{{\bf{R}}'_{\rm t}  = {\bf{R}}_{\rm t} }
\end{equation}
dramatically showing that it is entirely due to matter chirality through the DF mirror asymmetry. Such lateral momentum transfer testifies that the electron is subjected to a mechanical interaction orthogonal to $\pi$ which is geometrically analogous to, but physically different from, the  optical lateral force experienced by chiral nanoparticles \cite{Hayat,Shiii}.

\begin{figure}
\centering
\includegraphics[width = 1\linewidth]{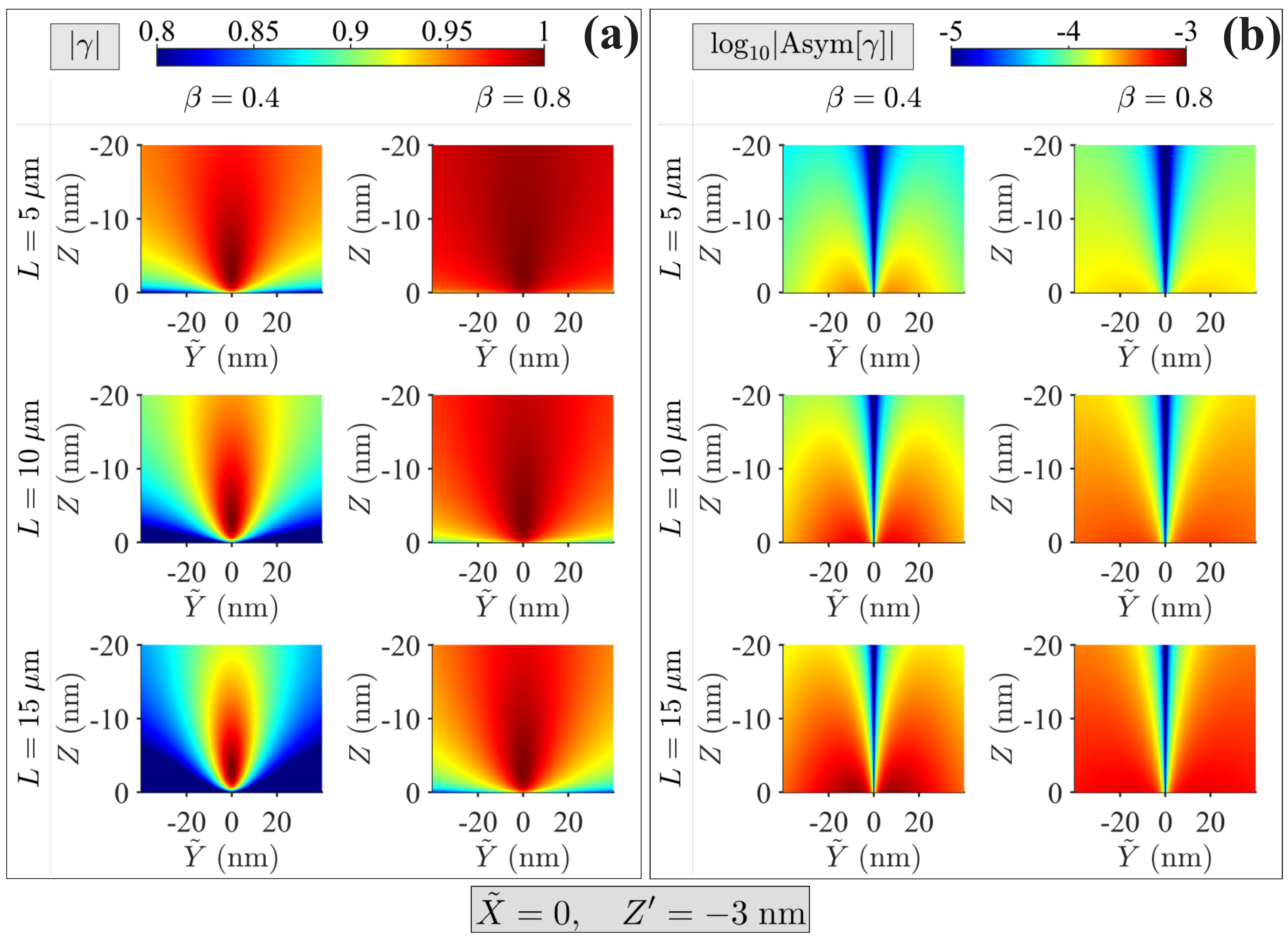}
\caption{Moduli of (a) the scattered electron decoherence factor $\gamma  ( {\tilde X,\tilde Y,Z, Z'} )$  and (b) its asymmetry degree ${\rm Asym}[\gamma] (\tilde X,\tilde Y, Z+Z')$ for $\tilde X=0$ and $Z' = -3 \; {\rm nm}$ and for various electron velocities $\beta = V/c$ (columns) and interaction lengths $L$ (rows).}
\label{Figure3}
\end{figure}

To discuss our general results in a realistic situation, we consider the setup sketched in Fig.2(a) where an aloof electron travels in vacuum ($\varepsilon_1 = 1$) with its velocity ${\bf V} = V {\bf e}_x$ parallel to a homogeneous chiral nanofilm of thickness $d = 50 \: {\rm nm}$ and deposited onto a substrate ($\varepsilon_2 = 1.48$), assuming infinite extension along the $y$ axis and finite length $L$ along the $x$ axis. We assume an elliptical Gaussian transverse profile for the initial wave function, $\phi _i \left( {Y,Z} \right) = \sqrt {\frac{2}{{\pi \sigma _y \sigma _z }}} e^{ - \left[ {\frac{{Y^2 }}{{\sigma _y^2 }} + \frac{{\left( {Z + d} \right)^2 }}{{\sigma _z^2 }}} \right]}$, of widths $\sigma_y$ and $\sigma_z$ and impact parameter $b$. In Fig.2(b) we plot the nanofilm permittivity $\varepsilon$ and chiral parameter $\kappa$ modelling typical chiral molecules with ultraviolet resonance at $3.54 \: {\rm eV}$ embedded in a dielectric matrix, with negligible magnetic response ($\mu =1$)  \cite{Chenn}. Such maximally symmetric setup has $\pi = zx$ as its reflection symmetry plane so that, from the comparison of Fig.1 and Fig.2(a), we get ${\cal R} = {\bf{e}}_x {\bf{e}}_x  - {\bf{e}}_y {\bf{e}}_y  + {\bf{e}}_z {\bf{e}}_z$, ${\bf{F}}_{\rm v}  = F_x {\bf{e}}_x$, ${\bf{F}}_\pi = F_z {\bf{e}}_z$, and ${\bf{F}}_{\rm n}  = F_y {\bf{e}}_y$; accordingly asymmetric effects show up along the $y$-axis which is the lateral direction. Planar geometry enables analytical evaluation of the relevant Green's tensor component along the electron velocity ${\cal G}_{\omega xx} \left( x-x',y-y',z+z'\right)$ and its reflected part in vacuum ($z<0$, $z'<0$) turns out to be mirror asymmetric (see Secs.S3.1 and S3.2 of Supplemental Material). Moreover, the antisymmetric part of ${\cal G}_{\omega xx}$ is produced by the reflection coefficients $R_{\rm SP}$, $R_{\rm PS}$ resulting from the $\rm S-P$ polarization coupling in turn triggered by matter chirality  \cite{Ciat1}.  The first of Eqs.(\ref{PhiDelta}) provides the phase factor $\Phi(Z)$, which is irrelevant in our symmetry analysis, whereas the second equation, after setting $\tilde X = X-X'$, $\tilde Y = Y-Y'$, $\tilde Z = Z+Z'$,  yields $\Delta ( {{\bf{R}},{\bf{R}}'} ) = \Delta _{\rm S} ( {\tilde X,\tilde Y,\tilde Z} ) + i\Delta _{\rm A} ( {\tilde X,\tilde Y,\tilde Z} )$ where the real $\Delta _{\rm S}$ and $\Delta _{\rm A}$ are proportional to the interaction length $L$ and are respectively symmetric and antisymmetric under the lateral reflection $\tilde Y \rightarrow - \tilde Y$, they stemming from the symmetric and antisymmetric parts of ${\cal G}_{\omega xx}$ (see Sec.S3.4 of Supplemental Material). All the electron mirror symmetry breaking effects are a consequence of $\Delta _{\rm A}$ which evidently vanishes in the achiral limit $\kappa \rightarrow 0$. In Fig.2(c) we plot $\Delta _{\rm S}$ and $\Delta _{\rm A}$ at $\tilde X = 0$ and $\tilde Z<0$ (vacuum) for the interaction length $L = 1 \: {\mu \rm m}$, for two electron velocities. Their relevant features are a consequence of the photon momentum $k_x =- \omega/V$ selected by the electron velocity and entailing the fully evanescent character of the field and the corresponing enhancement of the film chiral response. At lower electron velocities such evanescent character is more pronounced, since the photon momentum parallel to the film is $\sqrt { ( {\omega /V}  )^2  + k_y^2 }$, and accordingly both $\Delta _{\rm S}$ and $\Delta _{\rm A}$  show tighter spatial confinement and larger amplitudes.

\begin{figure*}
\centering
\includegraphics[width = 1\linewidth]{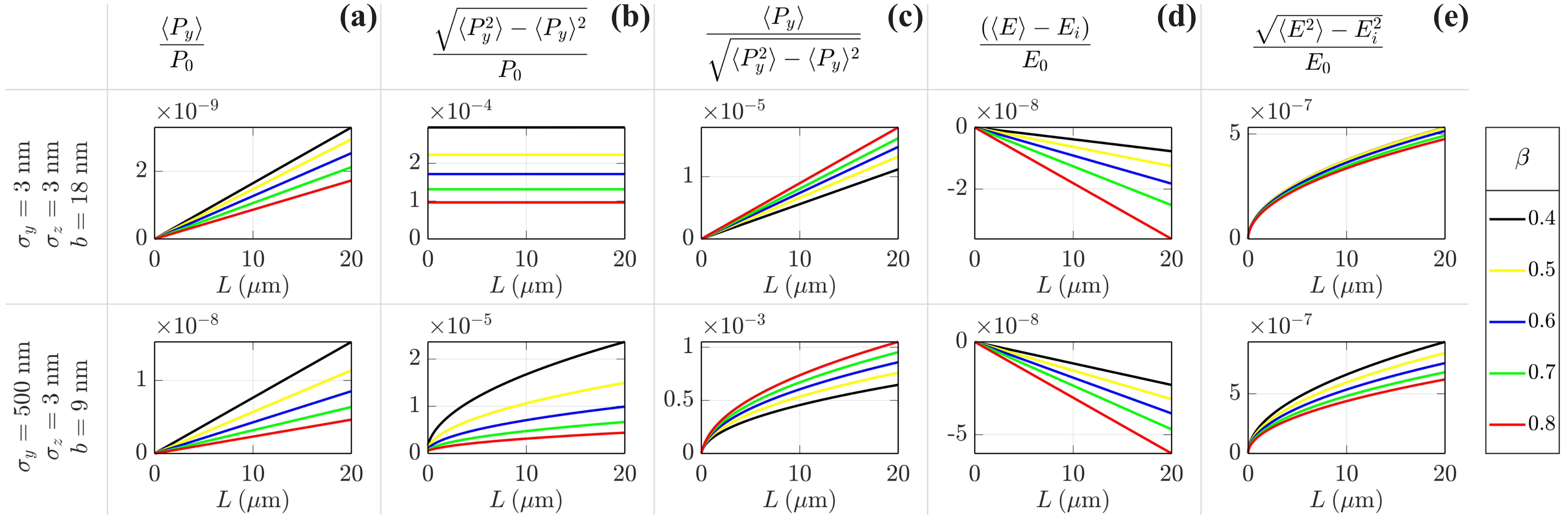}
\caption{Depence of the lateral momentum transfer and energy exchage on the interaction legth $L$ and electron velocity $\beta = V/c$ (colours) for two incident electron configurations (rows). Normalizations are performed with the reference momentum $P_0  = mV/\sqrt {1 - ( {V/c} )^2 }$ and energy $E_0 = VP_0$.}
\label{Figure4}
\end{figure*}

In Fig.3 we display the scattered electron DF $\gamma(\tilde X, \tilde Y , Z, Z')$ and its mirror symmetry breaking for various interaction lengths $L$ and two electron velocities, for $\tilde X =0$ and $Z' = - 3 \; {\rm nm}$. The modulus of $\gamma$ in Fig.3(a) turns out to be localized around the chosen reference $Z'$ and its spatial spread, for a fixed interaction length, is smaller at lower electron velocities, as a consequence of the evanescent mechanism ruling $\Delta$ discussed in Fig.2. In addition the spatial spot of $\gamma$ remarkably shrinks at larger interaction lengths, for a fixed electron velocity, an effect resulting from the non-perturbative character of the DF in Eq.(\ref{gam}) and testifying that first-order perturbation theory is not adequate in the here considered setup for interaction lengths of the order of a ten of microns (see Sec.2.6 of Supplemental Material). In Fig.3(b) we plot the modulus of the DF mirror asymmetry degree, defined as ${\rm Asym} \left[ {\gamma \left( {{\bf{R}},{\bf{R}}'} \right)} \right] = 2\frac{{\gamma  ( {{\bf{R}},{\bf{R}}'}  ) - \gamma  ( {{\cal R}{\bf{R}},{\cal R}{\bf{R}}'}  )}}{{\gamma  ( {{\bf{R}},{\bf{R}}'}  ) + \gamma  ( {{\cal R}{\bf{R}},{\cal R}{\bf{R}}'})}}$, which using Eq.(\ref{gam}) turns out to be 
\begin{equation}
{\rm Asym} \left[ \gamma  \right] = 2i\tan \Delta _{\rm A}.
\end{equation}
DF mirror asymmetry is again more pronouced at lower electron velocities owing to the above discussed behavior of $\Delta_{\rm A}$ and it remarkably increases at larger interaction lengths.

In Fig.4 we display the dependence of the overall phenomenology concerning lateral momentum transfer and energy exchange on the interaction length and the electron velocity, for two initial electron configurations: a tightly confined and distant electron with $(\sigma_y,\sigma_z,b) = (3, 3,18) \: {\rm nm}$ (first row) and a laterally wider and closer one with $(\sigma_y,\sigma_z,b) = (500, 3,9) \: {\rm nm}$ (second row). In Figs.4(a) and 4(b) we respectively plot mean value $\langle P_y \rangle$ and root-mean-square $\sqrt{{\left\langle {P_y^2 } \right\rangle  - \left\langle {P_y } \right\rangle ^2 }}$ (see Sec.S3.4 of Supplemental Material) of the electron lateral momentum (both normalized with the reference momentum $P_0  = mV/\sqrt {1 - ( {V/c} )^2 }$) showing that they increase both as the interaction length increases and as the electron velocity decreases, as a consequence of the properties of $\gamma$ discussed in Fig.3. However, the laterally wider and closer electron (second row) displays larger lateral momentum gain, since mirror symmetry breaking is more pronouced close to the film (see Fig.3(b)), and it has smaller momentum uncertainty. To quantifiy the efficiency of the lateral momentum transfer, in Fig.4(c) we plot the peak factor $\left\langle {P_y } \right\rangle /\sqrt {\left\langle {P_y^2 } \right\rangle  - \left\langle {P_y } \right\rangle ^2 }$ showing that the laterally wider and closer electron (second row) experiences a lateral momentum transfer two order of magnitudes more efficient than the tight confined and distant electron (first row). This is in agreement with the above general observation that the lateral smearing of the electron wave function rules the mirror symmetry breaking of its RDM. Remarkably, the lateral momentum transfer efficiency is globally larger at larger electron velocities. For completeness, in Figs.4(d) and 4(e) we respectively plot the loss mean value ${\left\langle E \right\rangle  - E_i }$ and root-mean-square $\sqrt {\left\langle {E^2 } \right\rangle  - \left\langle E \right\rangle ^2 }$  (see Sec.S3.5 of Supplemental Material) of the electron energy, both normalized with the reference energy $E_0  = VP_0$. The energy loss absolute value evidently increases as both the interaction length and electron velocity increase as expected. On the other hand the energy root-mean-square has the same behavior of the lateral momentum root-mean-square. Figures 5(a), (b), (d) and (e) provide the overall information that for the here considered relatively large interaction lengths (up to twenty microns), both the lateral momentum $P_y$ and the longitudinal one $P_x = E/V$ very slighlty depart from the reference momentum $P_0$ (for any electron velocity) and this self-consistently validates the above paraxial approximation without momentum recoil used to describe the electron dynamics.

In conclusion, we have shown that fast electrons with no handedness (i.e. carrying no orbital angular momentum) can experience the microscopic chirality of geometrically symmetric samples owing to the combination of electron lateral quantum smearing with the spatial asymmetry of the electromagnetic interaction. Matter chirality breaks mirror symmetry of electron quantum decoherence, an effect detectable through bi-prism electron holography \cite{Orcho,Potap} or in a two-slits interference experiment \cite{Frab1,Frab2}. This result is in agreement with the fact that the spatial coherence of inelastically scattered electrons is highly correlated to the optical properties of the sample. As a consequence, we predict that the electron lateral momentum distribution is asymmetric as well thus yielding a net lateral momentum transfer which can be detected through momentum-resolved energy loss spectroscopy \cite{Shekh}.

{\bf ACKNOWLEDGEMENTS} 

The author acknowledges PRIN 2017 PELM (grant number 20177PSCKT).

\end{document}


\title{Fast electrons interacting with chiral matter: mirror symmetry breaking of quantum decoherence and lateral momentum transfer: Supplemental Material}.

\author{A. Ciattoni$^1$}
\email{alessandro.ciattoni@spin.cnr.it}
\affiliation{$^1$CNR-SPIN, c/o Dip.to di Scienze Fisiche e Chimiche, Via Vetoio, 67100 Coppito (L'Aquila), Italy}
\date{\today}

\maketitle
\renewcommand\theequation{S\arabic{equation}}
\renewcommand\thefigure{S\arabic{figure}}
\renewcommand\thesection{S\arabic{section}}
\renewcommand\thesubsection{S\arabic{section}.\arabic{subsection}}
\count\footins = 1000

\tableofcontents

\newpage

\section{Macroscopic quantum electrodynamics in chiral media}

Macroscopic quantum electrodynamics (MQED) is a powerful tool providing the quantum description of the electromagnetic field in absorbing linear media \cite{Grune1,Dungg1} with any optical properties, encompassing spatial nonlocality and nonreciprocity \cite{Buhma1}. We here review the MQED formalism in inhomogenous chiral media characterized (in the  frequency domain $e^{-i \omega t}$) by the classical constitutive relations 
\begin{eqnarray} \label{consti}
 {\bf{D}}_\omega   &=& \varepsilon _0 \varepsilon    {\bf{E}}_\omega   - \frac{i}{c}\kappa    {\bf{H}}_\omega, \nonumber   \\ 
 {\bf{B}}_\omega   &=& \frac{i}{c}\kappa   {\bf{E}}_\omega   + \mu _0 \mu   {\bf{H}}_\omega, 
\end{eqnarray}
where $\varepsilon ({\bf r}, \omega)$ and $\mu  ({\bf r},\omega)$ are the relative permittivity and permeability and $\kappa ({\bf r}, \omega )$ is the Pasteur parameter controlling the coupling between electric and magnetic responses, i.e. the medium chirality; as required by causality, such  parameters satisfy the Kramers-Kronig dispersion relations. The details of the general quantization scheme are reported in Ref.\cite{Buhma1} and we here summarize the specific results pertaining chiral media.

\subsection{Green's tensor}   \label{GMAS}
The classical Green's tensor is among the basic tools of MQED and hence we briefly discuss here some of its properties. The classical field produced by a current density distribution ${\bf{J}}_{\omega \rm } \left( {{\bf{r}}} \right)$ in the presence of a chiral medium is
\begin{equation} \label{FieldEj}
{\bf{E}}_\omega  ( {\bf{r}} ) = i\omega \mu _0 \int {d^3 {\bf{r}'}} {\cal G}_\omega  ( {{\bf{r}},{\bf{r}}'} ) \:{\bf{J}}_{\omega \rm } \left( {{\bf{r}}'} \right)
\end{equation}
where the Green's tensor ${\cal G}_\omega({\bf r},{\bf r}')$ is the solution of the inhomogenous Helmholtz equation
\begin{equation} \label{Helm}
\left\{ {\nabla  \times \frac{1}{\mu }\nabla  \times  - \frac{{\omega ^2 }}{{c^2 }}\left( {\varepsilon  - \frac{{\kappa ^2 }}{\mu }} \right) + \frac{\omega }{c}\left[ {\left( {\nabla \frac{\kappa }{\mu }} \right) \times  + 2 \frac{{\kappa }}{\mu }\nabla  \times } \right]} \right\}{\cal G}_\omega  \left( {{\bf{r}},{\bf{r}}'} \right) = \delta \left( {{\bf{r}} - {\bf{r}}'} \right)I,
\end{equation}
with the boundary conditions ${\cal G}_\omega  \left( {{\bf{r}},{\bf{r}}'} \right) \to 0$ for ${\bf{r}},{\bf{r}}' \to \infty$. Here we have considered the general inhomogeneous situation where $\varepsilon$, $\mu$ and $\kappa$ are position dependent and it encompasses the case of piecewise homogeneous media.  A relevant property of the Green's tensor is that its analytic continuation ${\cal G}_\Omega  \left( {{\bf{r}},{\bf{r}}'} \right)$ to the $\Omega$ complex plane is analytic along the whole upper half plane ${\rm Im} \, \Omega  > 0$. In addition it satisfies the Schwarts reflection principle 
\begin{equation}
{{\cal G}_\Omega ^ *  \left( {{\bf{r}},{\bf{r}}'} \right) = {\cal G}_{ - \Omega ^* } \left( {{\bf{r}},{\bf{r}}'} \right)}
\end{equation}
and it displays the asymptotic behavior 
\begin{equation}
{\cal G}_\Omega  \left( {{\bf{r}},{\bf{r}}'} \right) \to  - \frac{{c^2 }}{{\Omega ^2 }}\delta \left( {{\bf{r}} - {\bf{r}}'} \right)
\end{equation} 
for $\Omega  \to \infty$. By using all these properties it is simple to show that the imaginary part of the Green's tensor satisfies the important integral relation
\begin{equation} \label{omImG}
\int\limits_0^{ + \infty } {d\omega } \, \omega \, {\mathop{\rm Im}\nolimits} \left[ {{\cal G}_\omega  \left( {{\bf{r}},{\bf{r}}'} \right)} \right] = \frac{\pi }{2}c^2 \delta \left( {{\bf{r}} - {\bf{r}}'} \right)I.
\end{equation}
Since chiral media are reciprocal, the Green's tensor satisfies the Onsager reciprocity relation 
\begin{equation} \label{Onsager}
{\cal G}_\omega ^{\rm T} \left( {{\bf{r}},{\bf{r}}'} \right) = {\cal G}_\omega  \left( {{\bf{r}}',{\bf{r}}} \right).
\end{equation}

\subsection{Field quantization}
Medium absorption forbids radiation to be regarded as an isolated system and hence, as suggested by the fluctuation-dissipation theorem, a noise current ${\bf{j}}_{\omega \rm }^N ({\bf r})$ is introduced to account for the medium degrees of freedom. Moreover, the noise current encodes the full radiation-medium state since its radiated field is provided by ${\bf{E}}_\omega  \left( {\bf{r}} \right) = i\omega \mu _0 \int {d^3 {\bf{r}}'} {\cal G}_\omega  \left( {{\bf{r}},{\bf{r}}'} \right){\bf{j}}_{\omega }^N \left( {{\bf{r}}'} \right)$ (see Eq.(\ref{FieldEj})). Quantization is performed by requiring that the electric and magnetic induction field operators satisfy the standard vacuum QED commutation relations and this leads to a radiation-medium description in terms of boson polaritonic excitations. Due to magnetoelectric coupling, both electric $(e)$ and magnetic $(m)$ contributions of the noise currents are involved and hence two field operators ${\bf{\hat f}}_{\omega e} \left( {\bf{r}} \right) = \hat f_{\omega ej} \left( {\bf{r}} \right){\bf{e}}_j $ and ${\bf{\hat f}}_{\omega m} \left( {\bf{r}} \right) = \hat f_{\omega mj} \left( {\bf{r}} \right){\bf{e}}_j$ are introduced in the Schr\"{o}dinger picture whose components satisfy the commutation relations
\begin{eqnarray} \label{commf}
 \left[ {\hat f_{\omega \lambda j} \left( {\bf{r}} \right),\hat f_{\omega '\lambda 'j'} \left( {{\bf{r}}'} \right)} \right] &=& 0, \nonumber \\ 
 \left[ {\hat f_{\omega \lambda j} \left( {\bf{r}} \right),\hat f_{\omega '\lambda 'j'}^\dag  \left( {{\bf{r}}'} \right)} \right] &=& \delta \left( {\omega  - \omega '} \right)\delta _{\lambda \lambda '} \delta _{jj'} \delta \left( {{\bf{r}} - {\bf{r}}'} \right)  
\end{eqnarray}
so that $\hat f_{\omega \lambda j}^\dag  \left( {\bf{r}} \right)$ is the creation operator of the $j$ component of a polaritonic excitation of frequency $\omega$ and kind $\lambda = e,m$ at the position $\bf r$. The electromagnetic Hamiltonian is
\begin{equation} \label{Hem}
\hat H_{em}  = \int\limits_0^\infty  {d\omega } \;\hbar \omega \sum\limits_\lambda  {\int {d^3 {\bf{r}}} \;{\bf{\hat f}}_{\omega \lambda }^\dag  \left( {\bf{r}} \right) \cdot {\bf{\hat f}}_{\omega \lambda } \left( {\bf{r}} \right)} 
\end{equation}
since it predicts the correct time evolution of the annihilation operators 
\begin{equation} \label{fomt}
e^{\frac{i}{\hbar }\hat H_{em} t} {\bf{\hat f}}_{\omega \lambda } e^{ - \frac{i}{\hbar }\hat H_{em} t}  = {\bf{\hat f}}_{\omega \lambda } e^{ - i\omega t}.
\end{equation} 
The electric field operator ${{\bf{\hat E}} = \int\limits_0^{ + \infty } {d\omega } \,  ( {{\bf{\hat E}}_\omega   + {\bf{\hat E}}_\omega ^\dag  } )}$ has positive frequency part 
\begin{equation} \label{Eom}
{\bf{\hat E}}_\omega  \left( {\bf{r}} \right) = \sum\limits_\lambda  {\int {d^3 {\bf{r}}'} {\cal G}_{\omega \lambda } \left( {{\bf{r}},{\bf{r}}'} \right){\bf{\hat f}}_{\omega \lambda } \left( {{\bf{r}}'} \right)} 
\end{equation} 
where the tensors ${\cal G}_{\omega e}$ and ${\cal G}_{\omega m}$ are obtained from the Green's tensor ${\cal G}_\omega$ by means of
\begin{equation}
\left( {\begin{array}{@{\mkern0mu} c @{\mkern0mu}}
   {{\cal G}_{\omega e} }  \\
   {{\cal G}_{\omega m} }  \\
\end{array}} \right)_{\left( {{\bf{r}},{\bf{r}}'} \right)}  =  - i\omega \mu _0 \sqrt {\frac{\hbar }{\pi }} \left( {\begin{array}{@{\mkern0mu} c c @{\mkern0mu}}
   {U_{ee} } & {U_{em} }  \\
   {U_{me} } & {U_{mm} }  \\
\end{array}} \right)_{\left( {\bf{r}} \right)} \left( {\begin{array}{@{\mkern0mu} c @{\mkern0mu}}
   {i\omega {\cal G}_\omega  }  \\
   {{\cal G}_\omega   \times \cev{\nabla }}  \\
\end{array}} \right)_{\left( {{\bf{r}},{\bf{r}}'} \right)} 
\end{equation}
with the matrix $U_{\lambda \lambda'}$ satisfying the equation
\begin{equation}
\left( {\begin{array}{@{\mkern0mu} c c @{\mkern0mu}} 
   {U_{ ee} } & {U_{ em} }  \\
   {U_{ me} } & {U_{ mm} }  \\
\end{array}} \right)^{* {\rm T}} \left( {\begin{array}{@{\mkern0mu} c c @{\mkern0mu}}
   {U_{ee} } & {U_{em} }  \\
   {U_{me} } & {U_{mm} }  \\
\end{array}} \right)  = \left( {\begin{array}{@{\mkern0mu} c c @{\mkern0mu}} 
   {\varepsilon _0 {\rm Im} \left( {\varepsilon  - \frac{{\kappa  ^2 }}{{\mu   }}} \right)} & {  i  
 {\sqrt {\frac{{\varepsilon _0 }}{{\mu _0 }}} }  {\mathop{\rm Im}\nolimits} \left( {\frac{{\kappa   }}{{\mu   }}} \right)}  \\
   { -i  {\sqrt {\frac{{\varepsilon _0 }}{{\mu _0 }}} }  {\mathop{\rm Im}\nolimits} \left( {\frac{{\kappa  }}{{\mu   }}} \right)} & { - \frac{1}{{\mu _0 }}{\rm Im}\left( {\frac{1}{{\mu   }}} \right)}  \\
\end{array}} \right).
\end{equation}
It is worth noting that the tensors  ${\cal G}_{\omega e}$ and ${\cal G}_{\omega m}$ satisfy the identity
\begin{equation} \label{GreenProp}
\sum\limits_\lambda  {\int {d^3 {\bf{s}}} \;{\cal G}_{\omega \lambda }  ( {{\bf{r}},{\bf{s}}}  ){\cal G}_{\omega \lambda }^{* \rm T}  ( {{\bf{r}}',{\bf{s}}}  )}  = \frac{{\,\hbar \omega ^2 }}{{\pi \varepsilon _0 c^2 }}{\mathop{\rm Im}\nolimits} \left[ {{\cal G}_\omega   ( {{\bf{r}},{\bf{r}}'} )} \right]
\end{equation}
which is very useful since, among other things, when combined with Eqs.(\ref{commf}) and (\ref{Eom}), it yields the relation
\begin{equation} \label{commutE}
\left[ {{\bf{F}} \cdot {\bf{\hat E}}_\omega  \left( {\bf{r}} \right),{\bf{F}}' \cdot {\bf{\hat E}}_{\omega '}^\dag  \left( {{\bf{r}}'} \right)} \right] = \delta \left( {\omega  - \omega '} \right)\frac{{\,\hbar \omega ^2 }}{{\pi \varepsilon _0 c^2 }}{\bf{F}} \cdot {\mathop{\rm Im}\nolimits} \left[ {{\cal G}_\omega  \left( {{\bf{r}},{\bf{r}}'} \right)} \right]{\bf{F}}'
\end{equation}
where $\bf{F}$ and $\bf{F}'$ are two vectors. Equation (\ref{commutE}) provides the commutator of electric field components at different points and different frequencies and it highlights the role played by the imaginary part of the Green's tensor.

\subsection{Vector and scalar potentials in the Coulomb gauge}
In order to deal with the electron-radiation coupling, the vector and scalar potentials, ${\bf{\hat A}} = \int\limits_0^{ + \infty } {d\omega } ( {{\bf{\hat A}}_\omega   + {\bf{\hat A}}_\omega ^\dag  } )$ and $\hat \phi  = \int\limits_0^{ + \infty } {d\omega } ( {\hat \phi _\omega   + \hat \phi _\omega ^\dag  } )$, are required. Their positive frequency parts are defined by the relation  ${\bf{\hat E}}_\omega   = i\omega {\bf{\hat A}}_\omega   - \nabla \hat \phi _\omega$ so that, after choosing the Coulomb gauge $\nabla  \cdot {\bf{\hat A}}_\omega  = 0$, the terms $i\omega {\bf{\hat A}}_\omega$ and $- \nabla \hat \phi _\omega$ can be identified with the transverse and longitudinal parts of the field ${\bf{\hat E}}_\omega$ so that
\begin{eqnarray} \label{AV0}
 {\bf{\hat A}}_\omega  \left( {\bf{r}} \right) &=& \frac{1}{{i\omega }}\int {d^3 {\bf{r}'}} \,\left[ {\frac{1}{{\left( {2\pi } \right)^3 }}\int {d^3 {\bf{k}}} \;e^{i{\bf{k}} \cdot  ( {{\bf{r}} - {\bf{r}}'}  )} \left( {I - \frac{{{\bf{kk}}}}{{k^2 }}} \right)} \right]  {\bf{\hat E}}_\omega  \left( {{\bf{r}}'} \right), \nonumber \\ 
 \hat \phi _\omega  \left( {\bf{r}} \right) &=& \int {d^3 {\bf{r}'}} \;\left[ {\frac{1}{{\left( {2\pi } \right)^3 }}\int {d^3 {\bf{k}}} \;e^{i{\bf{k}} \cdot  ( {{\bf{r}} - {\bf{r}}'}  )} \frac{{i{\bf{k}}}}{{k^2 }}} \right] \cdot {\bf{\hat E}}_\omega  \left( {{\bf{r}}'} \right).
\end{eqnarray}
We highlight that the choice of the Coulomb gauge is not incompatible with a possible spatial inhomogeneity of the medium (responsible for $\nabla  \cdot {\bf{\hat E}}_\omega   \ne 0$) since we are describing the electromagnetic field by means of both the vector and scalar potentials. This is made manifest by rewriting Eqs.(\ref{AV0}) as
\begin{eqnarray} \label{AV}
 {\bf{\hat A}}_\omega  \left( {\bf{r}} \right) &=& \frac{1}{{i\omega }}\left[ {{\bf{\hat E}}_\omega  \left( {\bf{r}} \right) + \nabla \int {d^3 {\bf{r}}'} \,\frac{{\nabla ' \cdot {\bf{\hat E}}_\omega  \left( {{\bf{r}}'} \right)}}{{4\pi \left| {{\bf{r}} - {\bf{r}}'} \right|}}} \right], \nonumber \\ 
 \hat \phi _\omega  \left( {\bf{r}} \right) &=& \int {d^3 {\bf{r}}'} \;\frac{{\nabla ' \cdot {\bf{\hat E}}_\omega  \left( {{\bf{r}}'} \right)}}{{4\pi \left| {{\bf{r}} - {\bf{r}}'} \right|}},
\end{eqnarray}
where we have used the Fourier integral representation of the Coulomb potential
\begin{equation}
\frac{1}{{4\pi r}} = \frac{1}{{\left( {2\pi } \right)^3 }}\int {d^3 {\bf{k}}} \frac{{e^{i{\bf{k}} \cdot {\bf{r}}} }}{{k^2 }}.
\end{equation}

\section{Interaction of a fast electron with a chiral medium}

\subsection{Interaction Hamiltonian}
The minimal coupling Hamiltonian describing a relativistic electron interacting with a chiral medium is
\begin{equation}
\hat H = c\sqrt {m^2 c^2  + [ {{\bf{\hat P}} + e{\bf{\hat A}} ( {{\bf{\hat R}}}  )} ]^2 }  - e\hat \phi  ( {{\bf{\hat R}}}  ) + \hat H_{em}
\end{equation}
where $m$ and $-e <0$ are the electron rest mass and charge, $\bf{\hat R}$ and $\bf{\hat P}$ are the electron position and momentum operators ($[ {{\bf{\hat R}},{\bf{\hat P}}} ] = i\hbar I$), $\bf{\hat A}$ and $\hat \phi$ are the above defined vector and scalar potentials and $\hat H_{em}$ is the electromagnetic Hamiltonian  defined of Eq.(\ref{Hem}). Since ${\bf{\hat A}} \cdot {\bf{\hat P}} - {\bf{\hat P}} \cdot {\bf{\hat A}} = i\hbar \nabla  \cdot {\bf{\hat A}}$, the trasversality of the vector potential in the Coulomb gauge implies that $( {{\bf{\hat P}} + e{\bf{\hat A}}} )^2  = \hat P^2  + 2e{\bf{\hat A}} \cdot {\bf{\hat P}} + e\hat A^2$ and hence, by exploiting the weakness of the electron-radiation coupling, the term $e^2 {\bf{\hat A}}^2$ can be neglected while the square root can be expanded up to the first order in the term $2e{\bf{\hat A}} \cdot {\bf{\hat P}}$ thus obtaining 
\begin{equation} \label{H2}
\hat H = c\sqrt {m^2 c^2  + \hat P^2 }  + e\frac{c}{{\sqrt {m^2 c^2  + \hat P^2 } }}{\bf{\hat A}} ( {{\bf{\hat R}}} ) \cdot {\bf{\hat P}} - e\hat \phi  ( {{\bf{\hat R}}} ) + \hat H_{em}.  
\end{equation}
To further simplify the Hamiltonian, we now perform the standard paraxial approximation, which is adequate to deal with highly collimated electron beams routinely used in TEM microscopes. Accordingly we consider electron states whose momentum distribution is strongly localized around a central momentum ${\bf P}_0$ ($| {{\bf{\hat P}} - {\bf{P}}_0 } | \ll P_0$) so that, by linearizing the first term of Eq.(\ref{H2}) around ${\bf P}_0$ and evaluating the second term at ${\bf P}_0$, after neglecting an irrelevant constant contribution, we get
\begin{equation} \label{H2-1}
\hat H = \underbrace {{\bf{V}} \cdot {\bf{\hat P}} + \hat H_{em} }_{\hat H_0 } + \underbrace {e\left[ {{\bf{V}} \cdot {\bf{\hat A}} ( {{\bf{\hat R}}}  ) - \hat \phi  ( {{\bf{\hat R}}}  )} \right]}_{\hat H_{int} }
\end{equation}
where ${\bf{V}} = c{\bf{P}}_0 /\sqrt {m^2 c^2  + P_0^2 }$. Due to the relation
\begin{equation} \label{Rt}
e^{\frac{i}{\hbar } ( {{\bf{V}} \cdot {\bf{\hat P}}}  )t} {\bf{\hat R}}e^{ - \frac{i}{\hbar } ( {{\bf{V}} \cdot {\bf{\hat P}}}  )t}  = {\bf{\hat R}} + {\bf{V}}t, 
\end{equation}
the vector $\bf V$ can be interpreted as the free electron velocity which is independent on momentum due to paraxial approximation. As highlighted by the curly brackets, the Hamiltonian has a main term ${\hat H_0}$ accounting for the bare electron-radiation system and a term ${\hat H_{int} }$ describing their coupling. By using Eq.(\ref{Hem}) and Eqs.(\ref{AV}) we get
\begin{align} \label{H3}
 \hat H_0  &= {\bf{V}} \cdot {\bf{\hat P}} + \int\limits_0^\infty  {d\omega } \;\hbar \omega \sum\limits_\lambda  {\int {d^3 {\bf{r}}} \;{\bf{\hat f}}_{\omega \lambda }^\dag  \left( {\bf{r}} \right) \cdot {\bf{\hat f}}_{\omega \lambda } \left( {\bf{r}} \right)}, \nonumber  \\ 
\hat H_{int}  &= \int\limits_0^{ + \infty } {d\omega } \frac{e}{{i\omega }}\int {d^3 {\bf{r}}} \left[ {{\bf{Q}}_\omega   ( {{\bf{\hat R}},{\bf{r}}}  ) \cdot {\bf{\hat E}}_\omega  \left( {\bf{r}} \right) - {\bf{Q}}_\omega ^*  ( {{\bf{\hat R}},{\bf{r}}}  ) \cdot {\bf{\hat E}}_\omega ^\dag  \left( {\bf{r}} \right)} \right] 
\end{align}
where
\begin{equation} \label{Q}
{\bf{Q}}_\omega  \left( {{\bf{R}},{\bf{r}}} \right) = \delta \left( {{\bf{R}} - {\bf{r}}} \right){\bf{V}} + \left( {{\bf{V}} \cdot \nabla _{\bf{R}}  - i\omega } \right)\frac{1}{{4\pi \left| {{\bf{R}} - {\bf{r}}} \right|}}\nabla _{\bf{r}}   \; . 
\end{equation}
The delta function contribution of ${\bf{Q}}_\omega$ provides the direct interaction between the electron and the radiation electric field whereas the seconds accounts for the coupling of the electron with the Coulomb field produced by the medium inhomogeneity.

\subsection{S-operator}
Let us now consider a typical scattering situation where a free electron in the distant past is made to interact with the chiral medium and it is subsequently collected by a detector in the far future. The time evolution of an initial electron-radiation state $\left| {\Psi _i } \right\rangle = \left| {\Psi} (- \infty ) \right\rangle$, in the interaction picture, is provided by $\left| {\Psi \left( t \right)} \right\rangle  = \hat U\left( t \right)\left| {\Psi _i } \right\rangle$ where the time evolution operator $\hat U\left( t \right)$ is the solution of the equation
\begin{equation} \label{Evol1}
i\hbar \frac{{d\hat U}}{{dt}} = \left( {e^{\frac{i}{\hbar }\hat H_0 t} \hat H_{int} e^{ - \frac{i}{\hbar }\hat H_0 t} } \right)\hat U
\end{equation}
with initial value $\hat U\left( { - \infty } \right) = 1$. Dynamics of fast electrons interacting with non magnetoelectric media has been investigated to some extent \cite{Kfir,River1,Hayun1} and we here extend those reults to chiral media. We start by using Eqs.(\ref{fomt}) and Eqs.(\ref{Rt}), to cast Eq.(\ref{Evol1}) as
\begin{equation} \label{Evol2}
\frac{{d\hat U}}{{dt}} = \left\{ {\int\limits_0^{ + \infty } {d\omega } \int {d^3 {\bf{r}}} \;\frac{e}{{\hbar \omega }}\left[ { - e^{ - i\omega t} {\bf{Q}}_\omega   ( {{\bf{\hat R}} + {\bf{V}}t,{\bf{r}}}  ) \cdot {\bf{\hat E}}_\omega  \left( {\bf{r}} \right) + \;e^{i\omega t} {\bf{Q}}_\omega ^*  ( {{\bf{\hat R}} + {\bf{V}}t,{\bf{r}}}  ) \cdot {\bf{\hat E}}_\omega ^\dag  \left( {\bf{r}} \right)} \right]} \right\}\hat U
\end{equation}
which we solve by means of the trial solution 
\begin{equation} \label{trial}
\hat U = e^{i\hat W}
\end{equation}
where
\begin{equation} \label{WWW}
\hat W = \Xi ( {{\bf{\hat R}},t} ) + \int\limits_0^{ + \infty } {d\omega } \int {d^3 {\bf{r}}} \;\left[ {{\bf{N}}_\omega  ( {{\bf{\hat R}},{\bf{r}},t} ) \cdot {\bf{\hat E}}_\omega  \left( {\bf{r}} \right) + \;{\bf{N}}_\omega ^* ( {{\bf{\hat R}},{\bf{r}},t} ) \cdot {\bf{\hat E}}_\omega ^\dag  \left( {\bf{r}} \right)} \right]
\end{equation}
and the unknown function $\Xi$ and vector ${\bf{N}}_\omega$ are such that $\Xi ( {{\bf{\hat R}}, -\infty} ) = 0$, ${\bf{N}}_\omega  ( {{\bf{\hat R}},{\bf{r}}, -\infty } ) = 0$, as required by the initial value of $\hat U$. In order to substitute Eq.(\ref{trial}) into Eq.(\ref{Evol2}), we use the Sneddon's formula for the derivative of the exponential of an operator \cite{Wilco1} 
\begin{equation} \label{derTrial}
\frac{d}{{dt}}e^{i\hat W}  = i\int\limits_0^1 {d \varphi \;e^{i\varphi \hat W} } \frac{{d\hat W}}{{dt}}e^{i\left( {1 - \varphi } \right)\hat W}.
\end{equation}
whose evaluation requires the analysis of the commutator of ${\hat W}$ and ${\frac{{d\hat W}}{{dt}}}$. From Eq.(\ref{commutE}), we easily get
\begin{equation} \label{commutNdN}
\left[ {\hat W,\frac{{d\hat W}}{{dt}}} \right] = \frac{{\,2i\hbar }}{{\pi \varepsilon _0 c^2 }} \int\limits_0^{ + \infty } {d\omega } \omega ^2 \int {d^3 {\bf{r}}}  \int {d^3 {\bf{r}'}} \; {\mathop{\rm Im}\nolimits} \left\{ {{\bf{N}}_\omega   ( {{\bf{\hat R}},{\bf{r}},t}  ) \cdot {\mathop{\rm Im}\nolimits} \left[ {{\cal G}_\omega  \left( {{\bf{r}},{\bf{r}}'} \right)} \right]\frac{{\partial {\bf{N}}_\omega ^*  ( {{\bf{\hat R}},{\bf{r}}',t}  )}}{{\partial t}}} \right\}
\end{equation}
whose independece on the polaritonic operators  ${\bf{\hat f}}_{\omega \lambda } ,{\bf{\hat f}}_{\omega \lambda }^\dag$ implies that it commutes with both ${\hat W}$ and ${\frac{{d\hat W}}{{dt}}}$. As a consequence, the relation $\left[ {e^{i\xi \hat W} ,\frac{{d\hat W}}{{dt}}} \right] = \left[ {\hat W,\frac{{d\hat W}}{{dt}}} \right]i\xi e^{i\xi \hat W}$ allows Eq.(\ref{derTrial}) to be written as 
\begin{equation}
\frac{d}{{dt}}e^{i\hat W}  = \left( { - \frac{1}{2}\left[ {\hat W,\frac{{d\hat W}}{{dt}}} \right] + i\frac{{d\hat W}}{{dt}}} \right)e^{i\hat W}
\end{equation}
which, once compared with Eq.(\ref{Evol2}), states that $e^{i\hat W}$ is the evolution operator if
\begin{eqnarray} \label{NXi}
 \frac{{\partial {\bf{N}}_\omega  }}{{\partial t}} &=& \frac{{ie}}{{\hbar \omega }}e^{ - i\omega t} {\bf{Q}}_\omega  ( {{\bf{\hat R}} + {\bf{V}}t,{\bf{r}}} ), \nonumber \\ 
 \frac{{\partial \Xi }}{{\partial t}} &=& \frac{1}{{\,2i}}\left[ {\hat W,\frac{{d\hat W}}{{dt}}} \right].
\end{eqnarray}
After noting that Eq.(\ref{Q}) yields
\begin{equation} 
e^{ - i\omega t} {\bf{Q}}_\omega  ( {{\bf{\hat R}} + {\bf{V}}t,{\bf{r}}} ) = e^{ - i\omega t} \delta ( {{\bf{\hat R}}   - {\bf{r}}} +{\bf{V}}t ) \; {\bf{V}} + \frac{d}{{dt}}\left[ {\frac{{e^{ - i\omega t} }}{{4\pi  | {{\bf{\hat R}} + {\bf{V}}t - {\bf{r}}}  |}}\nabla _{\bf{r}} } \right],
\end{equation}
the integration of the first of Eqs.(\ref{NXi}) is straighforward and we get
\begin{equation} \label{N}
{\bf{N}}_\omega ( {{\bf{\hat R}},{\bf{r}},t} )  = \frac{{ie}}{{\hbar \omega }}\left[ e^{i\frac{\omega }{V} ( {\hat R_{\rm v}  - r_{\rm v} }  )}  {\delta  ( {{\bf{\hat R}}_{\rm t}  - {\bf{r}}_{\rm t} }  )  \theta  ( {\hat R_{\rm v}   - r_{\rm v}  + Vt}  ) \; {\bf{u}}_{\rm v}  + \frac{{e^{ - i\omega t} }}{{4\pi  | {{\bf{\hat R}} + {\bf{V}}t - {\bf{r}}}  |}}\nabla _{\bf{r}} } \right]
\end{equation}
where $\theta(\xi)$ is the unit step function, ${\bf{u}}_{\rm v}  = {\bf{V}}/V$ is the electron velocity unit vector and we have introduced the decomposition ${\bf{F}} = {\bf{F}}_{\rm t}  + {\bf{F}}_{\rm v}$ of a vector ${\bf{F}}$ into its parts transverse (t) and parallel (v) to $\bf V$, i.e. 
\begin{eqnarray} \label{decomp}
 {\bf{F}}_{\rm t}  &=& {\bf{F}} - \left( {{\bf{u}}_{\rm v}  \cdot {\bf{F}}} \right){\bf{u}}_{\rm v}, \nonumber  \\ 
 {\bf{F}}_{\rm v}  &=& \left( {{\bf{u}}_{\rm v}  \cdot {\bf{F}}} \right){\bf{u}}_{\rm v}  \equiv F_{\rm v} {\bf{u}}_{\rm v}.
\end{eqnarray}
The second of Eqs.(\ref{NXi}) can also be integrated by using Eq.(\ref{commutNdN}) and Eq.(\ref{N}) and, after some tedious but straightforward algebra, we get
\begin{eqnarray} \label{Xi}
 \Xi  ( {{\bf{\hat R}},t}  ) &=& \frac{{4\alpha }}{c}
 {\mathop{\rm Im}\nolimits}  \int\limits_0^{ + \infty } {d\omega } \int\limits_{ - \infty }^{\hat R_{\rm v}  + Vt} {dr_{\rm v} } \int\limits_{ - \infty }^{r_{\rm v} } {dr'_{\rm v} } e^{i\frac{\omega }{V} ( {r_{\rm v}  - r'_{\rm v} } )}  {\mathop{\rm Im}\nolimits} \left[ {\bf{u}}_{\rm v}  \cdot {{\cal G}_\omega   ( {{\bf{\hat R}}_{\rm t}  + r_{\rm v} {\bf{u}}_{\rm v} ,{\bf{\hat R}}_{\rm t}  + r'_{\rm v} {\bf{u}}_{\rm v} }  )} {\bf{u}}_{\rm v}  \right] + \nonumber  \\ 
 &+& \frac{\alpha }{{\pi c}} {\mathop{\rm Im}\nolimits} \;\int\limits_0^{ + \infty } {d\omega } \;\int {d^3 {\bf{r}}} \int\limits_{ - \infty }^{\hat R_{\rm v}  + Vt} {dr'_{\rm v} } \frac{{e^{i\frac{\omega }{V} ( {\hat R_{\rm v}  + Vt - r'_{\rm v} }  )} }}{{ | {{\bf{\hat R}} + {\bf{V}}t - {\bf{r}}}  |}}\nabla _{\bf{r}}  \cdot {\mathop{\rm Im}\nolimits} \left[ {{\cal G}_\omega   ( {{\bf{r}},{\bf{\hat R}}_{\rm t}  + r'_{\rm v} {\bf{u}}_{\rm v} }  )} \right]{\bf{u}}_{\rm v}  + \nonumber \\ 
 &+& \frac{\alpha }{{4\pi ^2 c}} \;\int\limits_0^{ + \infty } {d\omega } \omega \int {d^3 {\bf{r}}} \;\int {d^3 {\bf{r} '}}\int\limits_{ - \infty }^t {d\tau } \frac{{\nabla _{\bf{r}}  \cdot {\mathop{\rm Im}\nolimits} \left[ {{\cal G}_\omega  \left( {{\bf{r}},{\bf{r}}'} \right)} \right]\mathord{\buildrel{\lower3pt\hbox{$\scriptscriptstyle\leftarrow$}} 
\over \nabla } _{{\bf{r}}'} }}{{ | {{\bf{\hat R}} + {\bf{V}}\tau  - {\bf{r}}}  |  | {{\bf{\hat R}} + {\bf{V}}\tau  - {\bf{r}}'}  |}},
\end{eqnarray}
where 
\begin{equation}
\alpha  = \frac{{e^2 }}{{4\pi \varepsilon _0 \hbar c}} \cong \frac{1}{{137}}
\end{equation}
is the fine structure constant. By using the integral relation of Eq.(\ref{omImG}) it is simple proving that the last contribution in Eq.(\ref{Xi}) can be written as
\begin{equation}
\frac{\alpha }{{4\pi ^2 c}}\;\int\limits_0^{ + \infty } {d\omega } \omega \int {d^3 {\bf{r}}} \;\int {d^3 {\bf{r}'}} \int\limits_{ - \infty }^t {d\tau } \frac{{\nabla _{\bf{r}}  \cdot {\mathop{\rm Im}\nolimits} \left[ {{\cal G}_\omega  \left( {{\bf{r}},{\bf{r}}'} \right)} \right]\mathord{\buildrel{\lower3pt\hbox{$\scriptscriptstyle\leftarrow$}} 
\over \nabla } _{{\bf{r}'}} }}{{\left| {{\bf{\hat R}} + {\bf{V}}\tau  - {\bf{r}}} \right|\left| {{\bf{\hat R}} + {\bf{V}}\tau  - {\bf{r}'}} \right|}} = \frac{{\alpha c}}{2}\;\int\limits_{ - \infty }^t {d\tau } \int\limits_0^\infty  {dr} \frac{1}{{r^2 }}
\end{equation}
which, although infinite, will be omitted since it provides an unobservable phase factor (being independent on $\bf{\hat R}$). By inserting Eq.(\ref{N}) and (\ref{Xi}) into Eq.(\ref{WWW}), we get the full time evolution operator $\hat U = e^{i\hat W}$ as
\begin{equation}
\hat U = \exp  \left[i {\Xi  ( {{\bf{\hat R}},t}  )}  \right]\exp \left\{ {i\int\limits_0^{ + \infty } {d\omega } \int {d^3 {\bf{r}}} \;\left[ {{\bf{N}}_\omega   ( {{\bf{\hat R}},{\bf{r}},t}  ) \cdot {\bf{\hat E}}_\omega   ( {\bf{r}}  ) + \;{\bf{N}}_\omega ^*  ( {{\bf{\hat R}},{\bf{r}},t}  ) \cdot {\bf{\hat E}}_\omega ^\dag   ( {\bf{r}}  )} \right]} \right\}
\end{equation}
and two observations are in order. First, if the medium inhomogeneity is weak, all the terms containing the gradient operator $\nabla_{\bf r}$ can be neglected since it operates on the electric field (or the Green's tensor) whose divergence  is vanishingly small together with the volume charge density. In this circumstance the obtained expression of $\hat U$ coincides with the one deduced in literature for non magnetoelectric media (apart from the expression of the electric field in Eq.(\ref{Eom}) which is different due to the  presence of both electric and magnetic polaritonic excitations). Second, in the limiting case $t \rightarrow { + \infty }$ we are mainly concerned here, the second contributions to Eq.(\ref{N}) and Eq.(\ref{Xi}) vanish since they explicitly contain the Coulomb potential evaluated at the infinitely far position of the electron, and accordingly the S-operator $\hat S = \hat U(+\infty)$ turns out to be
\begin{equation} \label{Sop1}
\hat S = e^{i\Phi ( {{\bf{\hat R}} } )} e^{\hat a^\dag  ( {{\bf{\hat R}}} ) - \hat a( {{\bf{\hat R}}} )}   
\end{equation}
where, after relabelling $r_{\rm v} \rightarrow q$ and  $r'_{\rm v} \rightarrow q'$, we have introduced the phase factor $\Phi$ and the operator $\hat a$ which, for a vector $\bf F$, are given by
\begin{align} \label{zca}
 \Phi ( {{\bf{F}} } ) &= \frac{{4\alpha }}{c}  \int\limits_0^{ + \infty } {d\omega } \int\limits_{ - \infty }^{ + \infty } {dq} \int\limits_{ - \infty }^q {dq'\,} \sin \left[\frac{\omega }{V}( {q - q'} )\right]  {\mathop{\rm Im}\nolimits} \left[ {\bf{u}}_{\rm v}  \cdot {{\cal G}_\omega  ( {{\bf{F}}_{\rm t}  + q{\bf{u}}_{\rm v} ,{\bf{F}}_{\rm t}  + q'{\bf{u}}_{\rm v} } )} {\bf{u}}_{\rm v} \right]  ,  \nonumber \\ 
\hat a( {{\bf{F}}} ) &= \int\limits_0^{ + \infty } {d\omega } \frac{e}{{\hbar \omega }}\int\limits_{ - \infty }^{ + \infty } {dq} \,e^{i\frac{\omega }{V}( {F_{\rm v}  - q} )} {\bf{u}}_{\rm v}  \cdot {\bf{\hat E}}_\omega  ( {{\bf{F}}_{\rm t}  + q{\bf{u}}_{\rm v} } ). 
\end{align}
Note that the phase factor $\Phi({\bf F})=\Phi({\bf F}_{\rm t})$ only depends on the transverse part ${\bf F}_{\rm t}$ of the vector $\bf F$. We stress that the obtained $S$-operator in Eq.(\ref{Sop1}) coincides with the one discussed in literature for non magnetoelectric media, and we have here extended its validity to chiral media in full generality. 

The commutator of ${\hat a}( {{\bf{\hat R}}}  )$ and ${\hat a}^\dag( {{\bf{\hat R}}}  )$ is easily evaluated by means of Eq.(\ref{commutE}) and it turns out to be 
\begin{equation} \label{COMaad}
\left[ {\hat a( {{\bf{\hat R}}} ),\hat a^\dag  ( {{\bf{\hat R}}} )} \right] = \Delta ( {{\bf{\hat R}},{\bf{\hat R}}} )
\end{equation}
where the have introduced the function of two vectors $\bf F$ and $\bf F'$
\begin{equation} \label{DELTAAA}
\Delta \left( {{\bf{F}},{\bf{F}}'} \right) = \frac{{4\alpha }}{c}\int\limits_0^\infty  {d\omega } e^{ - i\frac{\omega }{V}\left( {F_{\rm v}  - F'_{\rm v} } \right)} \int\limits_{ - \infty }^{ + \infty } {dq} \int\limits_{ - \infty }^{ + \infty } {dq'} \,e^{i\frac{\omega }{V}\left( {q - q'} \right)}  {\mathop{\rm Im}\nolimits} \left[ {\bf{u}}_{\rm v}  \cdot {{\cal G}_\omega  \left( {{\bf{F}}_{\rm t}  + q{\bf{u}}_{\rm v} ,{\bf{F}}'_{\rm t}  + q'{\bf{u}}_{\rm v} } \right)} {\bf{u}}_{\rm v}  \right]
\end{equation}
which plays a fundamental role in our analysis (see below). Note that the Onsager reciprocity of the Green's tensor (see Eq.(\ref{Onsager})) implies that 
\begin{equation}
\Delta \left( {{\bf{F}'},{\bf{F}}} \right) = \Delta ^* \left( {{\bf{F}},{\bf{F}'}} \right)
\end{equation}
and consequently $\Delta ( {{\bf{F}},{\bf{F}}} )$ a is real number and $\Delta ( {{\bf{\hat R}},{\bf{\hat R}}} )$ is an hermitian operator in agreement with the fact that, from Eq.(\ref{COMaad}), it is the commutator of two hermitian conjugated operators. Note also that the relation
\begin{equation} \label{longShift}
\Delta \left( {{\bf{F}},{\bf{F}}'} \right) = \Delta \left( {{\bf{F}}_{\rm t}  + \left( {F_{\rm v}  - F'_{\rm v} } \right){\bf{u}}_{\rm v} ,{\bf{F}}'_{\rm t} } \right)
\end{equation}
holds, so that $\Delta \left( {{\bf{F}},{\bf{F}}} \right) = \Delta \left( {{\bf{F}}_{\rm t} ,{\bf{F}}'_{\rm t} } \right)$ only depends on ${\bf{F}}_{\rm t}$. Since both ${\hat a}( {{\bf{\hat R}}}  )$ and ${\hat a}^\dag( {{\bf{\hat R}}}  )$ evidently commute with their commutator, we can resort to the Baker-Campbell-Hausdorff identity $e^{\hat X + \hat Y}  = e^{ - \frac{1}{2} [ {\hat X,\hat Y}  ]} e^{\hat X} e^{\hat Y}$ to cast Eq.(\ref{Sop1}) in the form
\begin{equation} \label{Sop2}
\hat S = e^{i\Phi ( {\bf{\hat R}}_{\rm t}  )} e^{ - \frac{1}{2}\Delta ( {\bf{\hat R}}_{\rm t},{\bf{\hat R}}_{\rm t} )} e^{\hat a^\dag  ( {{\bf{\hat R}}} )} e^{ - \hat a( {{\bf{\hat R}}} )}
\end{equation}
where we have made explicit the dependence of $\Phi$ and $\Delta$ (with equal arguments) on the transvere component $\bf{\hat R}_{\rm t}$ of the position operator.

\subsection{Electron reduced density matrix}

The S-operator obtained in the above section provides the full time evolution of the initial electron-radiation state $\left| {\Psi _i } \right\rangle$ toward the final state  $\left| \Psi  \right\rangle  = \hat S\left| {\Psi _i } \right\rangle$. Since radiation is not excited at first, the initial state we choose is 
\begin{equation} \label{Psii}
\left| {\Psi _i } \right\rangle  = \left| {\psi _i ,0} \right\rangle  \equiv \left| {\psi _i } \right\rangle  \otimes \left| 0 \right\rangle 
\end{equation}
where $\left| {\psi _i } \right\rangle$ is a free electron state of energy $E_i$ \cite{deAba}, i.e.
\begin{equation} \label{eigenH0}
{\bf{V}} \cdot {\bf{\hat P}}\left| {\psi _i } \right\rangle  = E_i \left| {\psi _i } \right\rangle
\end{equation}
and $\left| 0 \right\rangle$ is the vacuum radiation state definded by $\hat f_{\omega \lambda j} ( {\bf{r}} )\left| 0 \right\rangle  = 0$. Accordingly $\left| {\Psi _i } \right\rangle$ is an eigenstate of $\hat H_0$ with eigenvalue $E_i$. After applying the S-operator of Eq.(\ref{Sop2}) to such initial state and using the second of Eqs.(\ref{zca}) and Eq.(\ref{Eom}), we get
\begin{equation} \label{Psi1}
\left| \Psi  \right\rangle  = e^{i\Phi ( {{\bf{\hat R}}}_{\rm t} )} e^{ - \frac{1}{2}\Delta ( {{\bf{\hat R}}_{\rm t},{\bf{\hat R}}}_{\rm t} )} \sum\limits_{n = 0}^\infty  {\frac{1}{{n!}}} \left[ {\int {d\xi } \;T ( {{\bf{\hat R}},\xi } ) \hat f^\dag  \left( \xi  \right)} \right]^n \left| {\psi _i ,0} \right\rangle 
\end{equation}
where the exponential of $\hat a^\dag  ( {{\bf{\hat R}}}  )$ has been expanded, we have intoduced the compact index $\xi  = ( {\omega ,\lambda ,j,{\bf{r}}} )$ to label the elementary polaritonic excitation and, for notational convenience, we have set
\begin{eqnarray} \label{T}
\hat f\left( \xi  \right) &=& \hat f_{\omega \lambda j} \left( {\bf{r}} \right), \nonumber \\
\int {d\xi }  &=& \int {d\omega } \sum\limits_{\lambda j} {\int {d^3 {\bf{r}}} }, \nonumber \\
T( {{\bf{\hat R}},\xi } ) &=& \frac{e}{{\hbar \omega }}\int\limits_{ - \infty }^{ + \infty } {dq} \;e^{ - i\frac{\omega }{V}( {\hat R_{\rm v}  - q} )} {\bf{u}}_{\rm v}  \cdot {\cal G}_{\omega \lambda }^* ( {{\bf{\hat R}}_{\rm t}  + q{\bf{u}}_{\rm v} ,{\bf{r}}} ){\bf{e}}_j.
\end{eqnarray}
In order to analyze the final state $\left| \Psi  \right\rangle$, it is convenient to use the representation provided by the states
\begin{equation}
\left| {{\bf{R}},\xi _1  \ldots \xi _{\rm n} } \right\rangle  = \left| {\bf{R}} \right\rangle  \otimes \left[ {\frac{1}{{\sqrt {n!} }}\hat f^\dag  \left( {\xi _1 } \right) \ldots \hat f^\dag  \left( {\xi _{\rm n} } \right)\left| 0 \right\rangle } \right]
\end{equation}
where ${\left| {\bf{R}} \right\rangle }$ is the eigenvector of the electron position operator $\hat {\bf R}$ (i.e. ${\bf{\hat R}}\left| {\bf{R}} \right\rangle  = {\bf{R}}\left| {\bf{R}} \right\rangle$). The state $\left| {{\bf{R}},\xi _1  \ldots \xi _{\rm n} } \right\rangle$  corresponds to the electron at the position $\bf R$ and the radiation with $n$ quanta distributed on the $n$ polaritonic excitations ${\xi _1  \ldots \xi _{\rm n} }$. These states are an orthonormal basis of the tensor product of the electron Hilbert space and the radiation Fock space due to the relations
\begin{eqnarray}
\left\langle {{{\bf{R}},\xi _1  \ldots \xi _{\rm n} }}
 \mathrel{\left | {\vphantom {{{\bf{R}},\xi _1  \ldots \xi _{\rm n} } {{\bf{R}}',\xi '_1  \ldots \xi '_{n'} }}}
 \right. \kern-3pt}
 {{{\bf{R}}',\xi '_1  \ldots \xi '_{n'} }} \right\rangle  &=& \delta \left( {{\bf{R}} - {\bf{R}}'} \right)\frac{{\delta _{n,n'} }}{{n!}}\sum\limits_{\pi  \in S_{\rm n} } {\left[ {\delta \left( {\xi _1  - \xi '_{\pi \left( 1 \right)} } \right) \ldots \delta \left( {\xi _{\rm n}  - \xi '_{\pi \left( n \right)} } \right)} \right]} , \nonumber  \\ 
 \int {d^3 {\bf{R}}} \sum\limits_{n = 0}^\infty  {\int {d\xi _1  \cdots d\xi _{\rm n} } } \left| {{\bf{R}},\xi _1  \ldots \xi _{\rm n} } \right\rangle \left\langle {{\bf{R}},\xi _1  \ldots \xi _m } \right| &=& I, 
\end{eqnarray}
where the sum in the first equation spans the $n!$ permutation $\pi$ of the symmetric group $S_{\rm n}$ and 
\begin{equation}
\delta \left( {\xi  - \xi '} \right) = \delta \left( {\omega  - \omega '} \right)\delta _{\lambda \lambda '} \delta _{jj'} \delta \left( {{\bf{r}} - {\bf{r}}'} \right).
\end{equation}
Accordingly the state $\left| \Psi  \right\rangle$ of Eq.(\ref{Psi1}) can be written as
\begin{equation} \label{Psi2}
\left| \Psi  \right\rangle  = \int {d^3 {\bf{R}}} \sum\limits_{n = 0}^\infty  {\int {d\xi _1  \cdots d\xi _{\rm n} } } \left| {{\bf{R}},\xi _1  \ldots \xi _{\rm n} } \right\rangle \left\langle {{{\bf{R}},\xi _1  \ldots \xi _{\rm n} }}
 \mathrel{\left | {\vphantom {{{\bf{R}},\xi _1  \ldots \xi _{\rm n} } \Psi }}
 \right. \kern-3pt}
 {\Psi } \right\rangle 
\end{equation}
where
\begin{equation} \label{proj}
\left\langle {{{\bf{R}},\xi _1  \ldots \xi _{\rm n} }}
 \mathrel{\left | {\vphantom {{{\bf{R}},\xi _1  \ldots \xi _{\rm n} } \Psi }}
 \right. \kern-3pt}
 {\Psi } \right\rangle  = \psi _i ( {\bf{R}} ) e^{i\Phi ( {\bf{R}}_{\rm t} )} e^{ - \frac{1}{2}\Delta ( {{\bf{R}}_{\rm t},{\bf{R}}}_{\rm t} )} \frac{1}{{\sqrt {n!} }}T\left( {{\bf{R}},\xi _1 } \right) \ldots T\left( {{\bf{R}},\xi _{\rm n} } \right)
\end{equation}
and $\psi _i ( {\bf{R}} ) = \left\langle {{\bf{R}}} \mathrel{\left | {\vphantom {{\bf{R}} {\psi _i }}}  \right. \kern-3pt} {{\psi _i }} \right\rangle$ is the initial electron wavefunction. Equation (\ref{Psi2}) clearly reveals the entanglement between electron and radiation created by their interaction produced by the chiral medium. In the present work we are mainly interested in spatial effects produced by the chiral medium on the electron state, regardless of the radiation state. Accordingly, due to the entaglement, it is necessary to describe the electron in terms of its reduced density operator $\hat \rho _e$ obtained by taking the partial trace of the density operator $\hat \rho  = \left| \Psi  \right\rangle \left\langle \Psi  \right|$ over the radiation states. In the position representation which is here the most covenient one, we have
\begin{equation}
\hat \rho _e  = \int {d^3 } {\bf{R}}\int {d^3 } {\bf{R}}'\rho _e \left( {{\bf{R}},{\bf{R}}'} \right)\left| {\bf{R}} \right\rangle \left\langle {{\bf{R}}'} \right|
\end{equation}
where the reduced density matrix is
\begin{equation}
\rho _e ( {{\bf{R}},{\bf{R}}'} ) = \sum\limits_{n = 0}^\infty  {\int {d\xi _1  \ldots d\xi _{\rm n} } \left\langle {{{\bf{R}},\xi _1  \ldots \xi _{\rm n} }}
 \mathrel{\left | {\vphantom {{{\bf{R}},\xi _1  \ldots \xi _{\rm n} } \Psi }}
 \right. \kern-3pt}
 {\Psi } \right\rangle \left\langle {\Psi }
 \mathrel{\left | {\vphantom {\Psi  {{\bf{R}}',\xi _1  \ldots \xi _{\rm n} }}}
 \right. \kern-3pt}
 {{{\bf{R}}',\xi _1  \ldots \xi _{\rm n} }} \right\rangle }. 
\end{equation}
By using the projections of Eq.(\ref{proj}) we straightforwardly get

\begin{equation} \label{rhoe0}
\rho _e \left( {{\bf{R}},{\bf{R}}'} \right) = \psi _i \left( {\bf{R}} \right)\psi _i^* \left( {{\bf{R}}'} \right)e^{i\left[ {\Phi \left( {{\bf{R}}_{\rm t} } \right) - \Phi \left( {{\bf{R}}'_{\rm t} } \right)} \right]} e^{ - \frac{1}{2}\left[ {\Delta \left( {{\bf{R}}_{\rm t} ,{\bf{R}}_{\rm t} } \right) + \Delta \left( {{\bf{R}}'_{\rm t} ,{\bf{R}}'_{\rm t} } \right)} \right]} \sum\limits_{n = 0}^\infty  {\frac{1}{{n!}}\left[ {\int {d\xi } \, T\left( {{\bf{R}},\xi } \right)T^* \left( {{\bf{R}}',\xi } \right)} \right]^n }  
\end{equation}
which can be casted as
\begin{equation} \label{rhoe1}
\rho _e \left( {{\bf{R}},{\bf{R}}'} \right) = \psi _i \left( {\bf{R}} \right)\psi _i^* \left( {{\bf{R}}'} \right)e^{i\left[ {\Phi \left( {{\bf{R}}_{\rm t} } \right) - \Phi \left( {{\bf{R}}'_{\rm t} } \right)} \right] - \frac{1}{2}\left[ {\Delta \left( {{\bf{R}}_{\rm t} ,{\bf{R}}_{\rm t} } \right) + \Delta \left( {{\bf{R}}'_{\rm t} ,{\bf{R}}'_{\rm t} } \right)} \right]} e^{\Delta \left( {{\bf{R}},{\bf{R}}'} \right)} 
\end{equation}
where we have summed the power series and we have used the integral
\begin{equation}
\int {d\xi } \;T\left( {{\bf{R}},\xi } \right)   T^* \left( {{\bf{R}}',\xi } \right) = \Delta \left( {{\bf{R}},{\bf{R}}'} \right)
\end{equation}
which is a simple consequence of Eqs.(\ref{T}) and (\ref{commutE}). Since $\psi _i \left( {\bf{R}} \right)\psi _i^* \left( {{\bf{R}}'} \right)$ is the density matrix of the fully coherent initial electron state, it is convenient writing Eq.(\ref{rhoe1}) as
\begin{equation} \label{rhoe}
\rho _e \left( {{\bf{R}},{\bf{R}}'} \right) = \psi _i \left( {\bf{R}} \right)\psi _i^* \left( {{\bf{R}}'} \right)\gamma \left( {{\bf{R}},{\bf{R}}'} \right)
\end{equation}
where 
\begin{equation} \label{gammmm}
\gamma \left( {{\bf{R}},{\bf{R}}'} \right) = e^{i\left[ {\Phi \left( {{\bf{R}}_{\rm t} } \right) - \Phi \left( {{\bf{R}}'_{\rm t} } \right)} \right] - \frac{1}{2}\left[ {\Delta \left( {{\bf{R}}_{\rm t} ,{\bf{R}}_{\rm t} } \right) + \Delta \left( {{\bf{R}}'_{\rm t} ,{\bf{R}}'_{\rm t} } \right)} \right]} e^{\Delta \left( {{\bf{R}}_{\rm t}  + \left( {R_{\rm v}  - R'_{\rm v} } \right){\bf{u}}_{\rm v} ,{\bf{R}}'_{\rm t} } \right)} 
\end{equation}
quantifies the decoherence of the electron produced by its interaction with the chiral medium. Note that we have used Eq.(\ref{longShift}) to make explicit the dependence of $\Delta \left( {{\bf{R}},{\bf{R}}'} \right)$ on the difference of $R_{\rm v}$ and $R'_{\rm v}$ which is also true for $\gamma$, i.e.
\begin{equation}
\gamma \left( {{\bf{R}},{\bf{R}}'} \right) = \gamma \left( {{\bf{R}}_{\rm t}  + \left( {R_{\rm v}  - R'_{\rm v} } \right){\bf{u}}_{\rm v} ,{\bf{R}}'_{\rm t} } \right).
\end{equation}
Note that $\gamma \left( {{\bf{R}},{\bf{R}}} \right) =1$ so that $\rho _e \left( {{\bf{R}},{\bf{R}}} \right) = \left| {\psi _i \left( {\bf{R}} \right)} \right|^2$, in agreement with the normalization condition 
\begin{equation} \label{normaliz}
{\rm Tr} \left( {\hat \rho _e } \right) = 1.
\end{equation}

\subsection{Electron transverse momentum and energy distributions}
The reduced density matrix derived in the above section fully describes the state of the electron after its interaction with the chiral film. 
We here extract from it the electron energy and transverse momentum distribution which is routinely measured in standard electron microscopes (momentum resolved energy-loss spectroscopy). We start by considering the electron states $\left|  {{\bf{P}}_{\rm t} ,E}  \right\rangle$ of definite transverse momentum ${\bf P}_{\rm t}$ and energy $E$, whose wavefunctions are 
\begin{equation}
\left\langle {{\bf{R}}}
 \mathrel{\left | {\vphantom {{\bf{R}} { {{\bf{P}}_{\rm t} ,E} }}}
 \right. \kern-3pt}
 {{\bf{P}}_{\rm t} ,E} \right\rangle  = \frac{1}{{2\pi \hbar \sqrt \ell  }}e^{\frac{i}{\hbar }{\bf{P}}_{\rm t}  \cdot {\bf{R}}_{\rm t} } e^{i\frac{E}{{\hbar V}}R_{\rm v} } 
\end{equation}
where $\ell$ is the longitudinal quantization length and $E =  \frac{{\hbar V}}{\ell }2\pi m$, with $m$ integer. These states are an orthonormal basis since
\begin{eqnarray} \label{BASIS}
 \int {d^2 {\bf{P}}_{\rm t} } \sum\limits_E \left|  {{\bf{P}}_{\rm t} ,E}  \right\rangle \left\langle  {{\bf{P}}_{\rm t} ,E} \right|  &=& I, \nonumber \\ 
 \left\langle {{ {{\bf{P}}_{\rm t} ,E} }}
 \mathrel{\left | {\vphantom {{ {{\bf{P}}_{\rm t} ,E} } { {{\bf{P}}'_{\rm t} ,E'} }}}
 \right. \kern-\nulldelimiterspace}
 {{ {{\bf{P}}'_{\rm t} ,E'} }} \right\rangle  &=& \delta \left( {{\bf{P}}_{\rm t}  - {\bf{P}}'_{\rm t} } \right)\delta _{E,E'},
\end{eqnarray}
and the continuum limit is obtained at the end of the calculations by letting $\ell \rightarrow +\infty$ and by using the relation 
\begin{equation} \label{continuum}
\sum\limits_E {}  \to \frac{\ell }{{2\pi \hbar V}}\int {dE}.
\end{equation}
By inserting the completenss relation (first on Eqs.(\ref{BASIS})) into the reduced density matrix normalization condition (Eq.(\ref{normaliz})) we get
\begin{equation} \label{one0}
1 = \int {d^2 {\bf{P}}_{\rm t} } \sum\limits_E {} \frac{1}{{\left( {2\pi \hbar } \right)^2 \ell }}\int {d^3 {\bf{R}}} '\int {d^3 {\bf{R}}} e^{ - \frac{i}{\hbar }{\bf{P}}_{\rm t}  \cdot \left( {{\bf{R}}_{\rm t}  - {\bf{R}}'_{\rm t} } \right)} e^{ - i\frac{E}{{\hbar V}}\left( {R_{\rm v}  - R'_{\rm v} } \right)} \rho _e \left( {{\bf{R}},{\bf{R}}'} \right).
\end{equation}
Since the impinging electron has energy $E_i$, we set for its intial wavefunction 
\begin{equation} \label{ini}
\psi _i \left( {\bf{R}} \right) = \frac{1}{{\sqrt \ell  }}\phi _i \left( {{\bf{R}}_{\rm t} } \right)e^{i\frac{{E_i }}{{\hbar V}}R_{\rm v} } ,
\end{equation}
where $\phi _i \left( {{\bf{R}}_{\rm t} } \right)$ is an arbitrary transverse profile normalized as $\int {d^2 } {\bf{R}}_{\rm t} \left| {\phi _i \left( {\bf{R}} \right)} \right|^2  = 1$ and accordingly, by using Eq.(\ref{rhoe}) and Eq.(\ref{ini}), Eq.(\ref{one0}) can be written as
\begin{eqnarray} \label{one1}
1 &=& \int {d^2 {\bf{P}}_{\rm t} } \frac{1}{{\left( {2\pi \hbar } \right)^2 }}\int {d^2 {\bf{R}}_{\rm t} } \int {d^2 {\bf{R}}'_{\rm t} } e^{ - \frac{i}{\hbar }{\bf{P}}_{\rm t}  \cdot \left( {{\bf{R}}_{\rm t}  - {\bf{R}}'_{\rm t} } \right)} \phi _i \left( {{\bf{R}}_{\rm t} } \right)\phi _i^* \left( {{\bf{R}}'_{\rm t} } \right) \times \nonumber \\
&\times& \sum \limits_E {\frac{1}{{\ell ^2 }}\int\limits_{ - \ell /2}^{\ell /2} {dR_{\rm v} } \int\limits_{ - \ell /2}^{\ell /2} {dR'_{\rm v} } e^{ - i\frac{{E - E_i }}{{\hbar V}}\left( {R_{\rm v}  - R'_{\rm v} } \right)} \gamma \left( {{\bf{R}}_{\rm t}  + \left( {R_{\rm v}  - R'_{\rm v} } \right){\bf{u}}_{\rm v} ,{\bf{R}}'_{\rm t} } \right)} .
\end{eqnarray}
After introducing the variable $Q = R_{\rm v}  - R'_{\rm v}$ and noting that the relation  
\begin{equation}
\int\limits_{ - \ell /2}^{\ell /2} {dR_{\rm v} } \int\limits_{ - \ell /2}^{\ell /2} {dR'_{\rm v} } \;F\left( {R_{\rm v}  - R'_{\rm v} } \right) = \int\limits_{ - \ell }^\ell  {dQ} \left( {\ell  - \left| Q \right|} \right)\;F\left( Q \right),
\end{equation}
holds for any function $F\left( {R_{\rm v}  - R'_{\rm v} } \right)$, the continuum limit of Eq.(\ref{one1}) can easily be  evaluated with the help of Eq.(\ref{continuum}) and it is
\begin{equation}
1 = \int {d^2 {\bf{P}}_{\rm t} } \int {dE} \frac{{d \mathscr{P}}}{{d^2 {\bf{P}}_{\rm t} dE}}
\end{equation}
where 
\begin{equation}
\frac{{d \mathscr{P}}}{{d^2 {\bf{P}}_{\rm t} dE}} = \frac{1}{{\left( {2\pi \hbar } \right)^3 V}}\int {d^2 {\bf{R}}_{\rm t} } \int {d^2 {\bf{R}}'_{\rm t} } \int {dQ} e^{ - \frac{i}{\hbar }{\bf{P}}_{\rm t}  \cdot \left( {{\bf{R}}_{\rm t}  - {\bf{R}}'_{\rm t} } \right)} e^{ - i\frac{{E - E_i }}{{\hbar V}}Q} \phi _i \left( {{\bf{R}}_{\rm t} } \right)\phi _i^* \left( {{\bf{R}}'_{\rm t} } \right)\gamma \left( {{\bf{R}}_{\rm t}  + Q{\bf{u}}_{\rm v} ,{\bf{R}}'_{\rm t} } \right)
\end{equation}
is the transverse momentum and energy probability distribution of the electron after the interaction with the chiral film. By integrating this expression over $E$ and over ${\bf{P}}_{\rm t}$, we respectively get the transverse momentum and energy probability distributions
\begin{eqnarray} \label{dP3}
 \frac{{d \mathscr{P}}}{{d^2 {\bf{P}}_{\rm t} }} &=& \frac{1}{{\left( {2\pi \hbar } \right)^2 }}\int {d^2 {\bf{R}}_{\rm t} } \int {d^2 {\bf{R}}'_{\rm t} } \; e^{ - \frac{i}{\hbar }{\bf{P}}_{\rm t}  \cdot \left( {{\bf{R}}_{\rm t}  - {\bf{R}}'_{\rm t} } \right)} \phi _i \left( {{\bf{R}}_{\rm t} } \right)\phi _i^* \left( {{\bf{R}}'_{\rm t} } \right)\gamma \left( {{\bf{R}}_{\rm t} ,{\bf{R}}'_{\rm t} } \right), \nonumber \\ 
\frac{{d \mathscr{P}}}{{dE}} &=& \frac{1}{{2\pi \hbar V}}\int {d^2 {\bf{R}}_{\rm t} } \int {dQ} \; e^{ - i\frac{{E - E_i }}{{\hbar V}}Q} \left| {\phi _i \left( {{\bf{R}}_{\rm t} } \right)} \right|^2 \gamma \left( {{\bf{R}}_{\rm t}  + Q{\bf{u}}_{\rm v} ,{\bf{R}}_{\rm t} } \right).
\end{eqnarray}
It is crucial here evaluating the mean value and variance of both transverse momentum and energy. We start by decomposing transverse position and momentum into their orthogonal components, i.e. ${\bf{R}}_{\rm t}  = R_{ 1} {\bf{u}}_1  + R_{2} {\bf{u}}_2$ and ${\bf{P}}_{\rm t}  = P_{1} {\bf{u}}_1  + P_{2} {\bf{u}}_2$, where ${\bf{u}}_1$ and ${\bf{u}}_2$ are orthogonal unit vectors spanning the transverse plane and we consider the $n$-th order moments 
\begin{eqnarray}
\left\langle {P_{j}^n } \right\rangle  = \int {d^2 {\bf{P}}_{\rm t} } \frac{{d \mathscr{P}}}{{d^2 {\bf{P}}_{\rm t} }}P_{j}^n, \nonumber \\
\left\langle {E^n } \right\rangle  = \int {dE} \frac{{d \mathscr{P}}}{{dE}}E^n.
\end{eqnarray}
By using the distributions of Eqs.(\ref{dP3}) we get 
\begin{eqnarray} \label{mom0}
 \left\langle {P_{j}^n } \right\rangle  &=& \int {d^2 {\bf{R}}_{\rm t} } \int {d^2 {\bf{R}}'_{\rm t} } \left[ {\left( { - \frac{\hbar }{i}\frac{\partial }{{\partial R_{j} }}} \right)^n \delta \left( {{\bf{R}}_{\rm t}  - {\bf{R}}'_{\rm t} } \right)} \right]\phi _i \left( {{\bf{R}}_{\rm t} } \right)\phi _i^* \left( {{\bf{R}}'_{\rm t} } \right)\gamma \left( {{\bf{R}}_{\rm t} ,{\bf{R}}'_{\rm t} } \right), \nonumber  \\ 
 \left\langle {E^n } \right\rangle  &=& \int {d^2 {\bf{R}}_{\rm t} } \int {dQ} \left[ {\left( { - V\frac{\hbar }{i}\frac{\partial }{{\partial Q}}} \right)^n \delta \left( Q \right)} \right] e^{i\frac{{E_i }}{{\hbar V}}Q} \left| {\phi _i \left( {{\bf{R}}_{\rm t} } \right)} \right|^2 \gamma \left( {{\bf{R}}_{\rm t}  + Q{\bf{u}}_{\rm v} ,{\bf{R}}_{\rm t} } \right) 
\end{eqnarray}
which can be written, after integrating by parts $n$-times and performing the integration over ${\bf{R}}'_{\rm t}$ and $Q$ with the help of the delta functions, as
\begin{eqnarray} \label{mom1}
 \left\langle {P_j^n } \right\rangle  &=& \int {d^2 {\bf{R}}_{\rm t} } \phi _i^* \left( {{\bf{R}}_{\rm t} } \right)\left\{ {\hat P_j^n \left[ {\phi _i \left( {{\bf{R}}_{\rm t} } \right)\gamma \left( {{\bf{R}}_{\rm t} ,{\bf{R}}'_{\rm t} } \right)} \right]} \right\}_{{\bf{R}}'_{\rm t}  = {\bf{R}}_{\rm t} }, \nonumber \\ 
 \left\langle {E^n } \right\rangle  &=& \int {d^2 {\bf{R}}_{\rm t} } \left| {\phi _i \left( {{\bf{R}}_{\rm t} } \right)} \right|^2 \left\{ {\hat E^n \left[ {e^{i\frac{{E_i }}{{\hbar V}}Q} \gamma \left( {{\bf{R}}_{\rm t}  + Q{\bf{u}}_{\rm v} ,{\bf{R}}_{\rm t} } \right)} \right]} \right\}_{Q = 0},
\end{eqnarray}
where we have introduced the transverse momentum $\hat P_j  = \frac{\hbar }{i}\frac{\partial }{{\partial R_j }}$ and energy $\hat E = V\frac{\hbar }{i}\frac{\partial }{{\partial Q}}$ operators. For $n=1$ these relations readly provide the transverse momentum and energy mean values
\begin{eqnarray} \label{meanvalues}
 \left\langle {P_j } \right\rangle  &=& \int {d^2 {\bf{R}}_{\rm t} } \phi _i^* \hat P_j \phi _i  + \int {d^2 {\bf{R}}_{\rm t} } \left| {\phi _i } \right|^2 \left( {\hat P_j \Theta } \right)_{\scriptstyle Q = 0 \hfill \atop 
  \scriptstyle {\bf{R}}'_{\rm t}  = {\bf{R}}_{\rm t}  \hfill}, \nonumber  \\ 
 \left\langle E \right\rangle  &=& E_i  + \int {d^2 {\bf{R}}_{\rm t} } \left| {\phi _i } \right|^2 \left( {\hat E\Theta } \right)_{\scriptstyle Q = 0 \hfill \atop 
  \scriptstyle {\bf{R}}'_{\rm t}  = {\bf{R}}_{\rm t}  \hfill},
\end{eqnarray}
where
\begin{equation} \label{THETA}
\Theta \left( {{\bf{R}}_{\rm t}  + Q{\bf{u}}_{\rm v} ,{\bf{R}}'_{\rm t} } \right) = \log \gamma \left( {{\bf{R}}_{\rm t}  + Q{\bf{u}}_{\rm v} ,{\bf{R}}'_{\rm t} } \right).
\end{equation}
Since $\int {d^2 {\bf{R}}_{\rm t} } \phi _i^* \hat P_j \phi _i$ is the transverse momentum mean value of the initial electron state, Eqs. (\ref{meanvalues}) remarkably show that both expectation values are the sum of the initial contribution plus a term resulting from the interaction with the chiral medium which is the average on the initial electron state of suitable  derivatives of $\Phi$ and  $\Delta$. Analagously, Eqs.(\ref{mom1}) for $n=2$ enable the evaluation of the variances which, after some algebra, turn out to be
\begin{eqnarray}   \label{variances}
 \left\langle {P_j^2 } \right\rangle  - \left\langle {P_j } \right\rangle ^2  &=& \left[ {\int {d^2 {\bf{R}}_{\rm t} } \phi _i^* \hat P_j^2 \phi _i  - \left( {\int {d^2 {\bf{R}}_{\rm t} } \phi _i^* \hat P_j \phi _i } \right)^2 } \right] +  	\nonumber \\
 &+& 2 \left\{ {\int {d^2 {\bf{R}}_{\rm t} } \left( {\phi _i^* \hat P_j \phi _i } \right)\left( {\hat P_j \Theta } \right)_{\scriptstyle Q = 0 \hfill \atop 
  \scriptstyle {\bf{R}}'_{\rm t}  = {\bf{R}}_{\rm t}  \hfill} }
 - \left( \int {d^2 {\bf{R}}_{\rm t} } \phi _i^* \hat P_j \phi _i \right) \left[ \int {d^2 {\bf{R}}_{\rm t} } \left| {\phi _i } \right|^2 \left( {\hat P_j \Theta } \right)_{\scriptstyle Q = 0 \hfill \atop 
  \scriptstyle {\bf{R}}'_{\rm t}  = {\bf{R}}_{\rm t}  \hfill} \right]  \right\} + \nonumber \\
  &+& \left\{ {\int {d^2 {\bf{R}}_{\rm t} } \left| {\phi _i } \right|^2 \left[ {\left( {\hat P_j \Theta } \right)^2  + \hat P_j^2 \Theta } \right]_{\scriptstyle Q = 0 \hfill \atop 
  \scriptstyle {\bf{R}}'_{\rm t}  = {\bf{R}}_{\rm t}  \hfill}  - \left[ {\int {d^2 {\bf{R}}_{\rm t} } \left| {\phi _i } \right|^2 \left( {\hat P_j \Theta } \right)_{\scriptstyle Q = 0 \hfill \atop 
  \scriptstyle {\bf{R}}'_{\rm t}  = {\bf{R}}_{\rm t}  \hfill} } \right]^2 } \right\}, \nonumber \\ 
 \left\langle {E^2 } \right\rangle  - \left\langle E \right\rangle ^2  &=& \int {d^2 {\bf{R}}_{\rm t} } \left| {\phi _i } \right|^2 \left[ {\left( {\hat E\Theta } \right)^2  + \hat E^2 \Theta } \right]_{\scriptstyle Q = 0 \hfill \atop 
  \scriptstyle {\bf{R}}'_{\rm t}  = {\bf{R}}_{\rm t}  \hfill}  - \left[ {\int {d^2 {\bf{R}}_{\rm t} } \left| {\phi _i } \right|^2 \left( {\hat E\Theta } \right)_{\scriptstyle Q = 0 \hfill \atop 
  \scriptstyle {\bf{R}}'_{\rm t}  = {\bf{R}}_{\rm t}  \hfill} } \right]^2.
\end{eqnarray}
From the first of Eqs.(\ref{variances}), the  transverse momentum variance is the sum of the transverse momentum variance of the intitial electron state (first term in square brackets) plus two contributions produced by the interaction of the electron with the chiral medium (second and third term in curly brackets). On the other hand, from the second of Eqs.(\ref{variances}), the energy variance has no initial electron contribution as expected since the impinging electron has definite energy $E_i$.

\begin{figure}
\centering
\includegraphics[width = 0.5\linewidth]{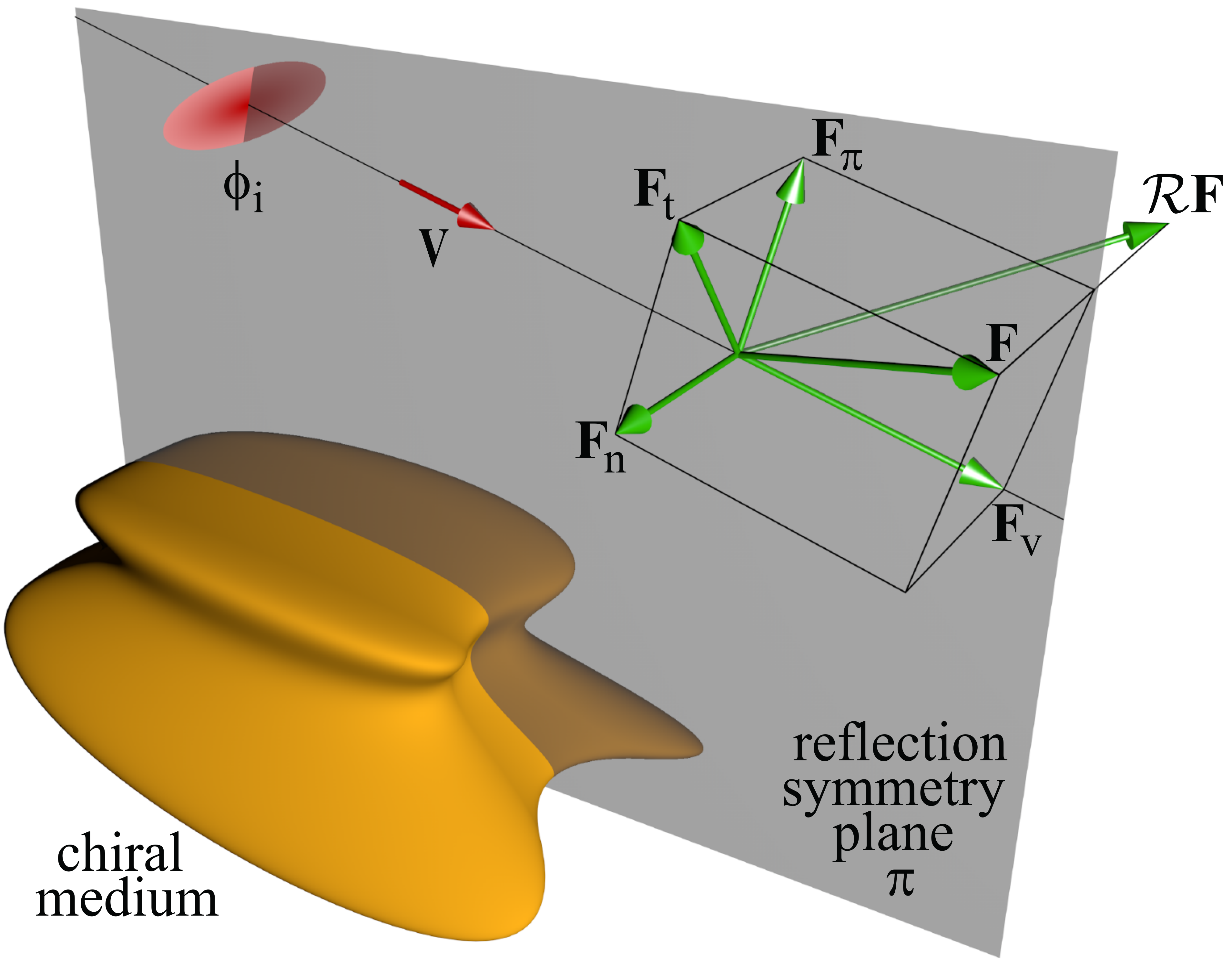}
\caption{Sketch of the electron-chiral medium geometry suitable for investigating mirror symmetry breaking effects. The plane $\pi$ is a reflection symmetry plane of the chiral medium and $\cal R$ is the geometrical reflection through $\pi$ which transform a vector $\bf F$ into its mirror image ${\cal R} {\bf F}$. The electron velocity $\bf V$ lies on $\pi$ and the electron initial wavefunction transverse profile $\phi_i$ is left invariant by $\cal R$. A vector admits the deconposition ${\bf{F}} = {\bf{F}}_{\rm t} + {\bf{F}}_{\rm v}$, into its parts transverse ($\rm t$) and parallel ($\rm v$) to $\bf V$ (see Eqs.(\ref{decomp})). In turn the transverse part admits the decomposition ${\bf{F}}_{\rm t}  = {\bf{F}}_{\pi}  + {\bf{F}}_{\rm n}$ into its parts parallel ($\pi$) and orthogonal ($\rm n$) to $\pi$. 
}
\label{FigureS1}
\end{figure}

\subsection{Mirror asymmetry of electron decoherence and lateral momentum transfer}
The obtained expressions of the reduced density matrix, transverse momentum and energy distributions are rather general with regard to their dependence on the Green's tensor, they formally holding also for achiral media. On the other hand, medium chirality has impact on the Green's tensor behavior under spatial reflections, a topic we here discuss as a prelude to the investigation of the mirror symmetry breaking of the electron quantum properties. To this end we consider the geometric spatial reflection ${\bf{r}}^{\rm M} = {\cal R}{\bf{r}}$ through an arbitrary mirror plane $\pi$ where ${\cal R} = I - 2 {\bf{nn}}$ is the reflection tensor and $\bf n$ is the unit vector normal to $\pi$. In order to avoid geometrical mirror asymmetries, we hereafter focus on a geometrically mirror symmetric medium for which 
\begin{eqnarray} \label{MirrorSymmMedium}
\varepsilon \left( {{\cal R}{\bf{r}}} ,\omega \right) &=& \varepsilon \left( {\bf{r}},\omega \right), \nonumber \\
\mu \left( {{\cal R}{\bf{r}}},\omega \right) &=& \mu \left( {\bf{r}},\omega \right), \nonumber \\
\kappa \left( {{\cal R}{\bf{r}}},\omega \right) &=& \kappa \left( {\bf{r}},\omega \right).
\end{eqnarray}
Since the electric field and the current densities are polar vectors, their mirror immages are ${\bf{E}}_\omega ^{\rm M} \left( {\bf{r}} \right) = {\cal R}{\bf{E}}_\omega  \left( {{\cal R}{\bf{r}}} \right)$ and ${\bf{j}}_\omega ^{\rm M} \left( {\bf{r}} \right) = {\cal R}{\bf{j}}_\omega  \left( {{\cal R}{\bf{r}}} \right)$ so that Eq.(\ref{FieldEj}) implies that the mirror image of the Green's tensor is
\begin{equation}
{\cal G}_\omega ^{\rm M} \left( {{\bf{r}},{\bf{r}}'} \right) = {\cal R}{\cal G}_\omega  \left( {{\cal R}{\bf{r}},{\cal R}{\bf{r}}'} \right){\cal R}.
\end{equation}
Now it is simple showing that ${\cal G}_\omega ^{\rm M} \left( {{\bf{r}},{\bf{r}}'} \right)$ satisfies the equation 
\begin{equation} \label{HelmM}
\left\{ {\nabla  \times \frac{1}{\mu }\nabla  \times  - \frac{{\omega ^2 }}{{c^2 }}\left( {\varepsilon  - \frac{{\kappa ^2 }}{\mu }} \right) + \frac{\omega }{c}\left[ {\left( {\nabla \frac{-\kappa }{\mu }} \right) \times  + 2 \frac{{-\kappa }}{\mu }\nabla  \times } \right]} \right\}{\cal G}_\omega ^{\rm M} \left( {{\bf{r}},{\bf{r}}'} \right) = \delta \left( {{\bf{r}} - {\bf{r}}'} \right)I
\end{equation}
which is identical to Eq.(\ref{Helm}) except for the sign switch of the chirality parameter $\kappa$. In other words, ${\cal G}_\omega ^{\rm M} \left( {{\bf{r}},{\bf{r}}'} \right)$ is the Green's tensor of the opposite enantiomeric medium which is the mirror image of the original one (since microscopically each chiral molecule is reflected into its opposite enantiomer). This observation is a mere restatement of reflection invariance (or mirror symmetry) of classical electrodynamics which states that if a complete experiment (i.e. field, sources and matter) is subjected to mirror reflection, the resulting experiment should, in principle, be realizable. What is remarkable here is that reflection invariance additionally entails the mirror asymmetry of the Green's tensor of a chiral medium, i.e. ${\cal G}_\omega ^{\rm M}  \left( {{\bf{r}},{\bf{r}}'} \right) \ne {\cal G}_\omega  \left( {{\bf{r}},{\bf{r}}'} \right)$. To prove this we note that the assumption ${\cal G}_\omega ^{\rm M} \left( {{\bf{r}},{\bf{r}}'} \right) = {\cal G}_\omega  \left( {{\bf{r}},{\bf{r}}'} \right)$ provides the equations
\begin{eqnarray} \label{AbsMirrorG}
 \nabla  \times \left[ {\frac{1}{\mu }\nabla  \times {\cal G}_\omega  \left( {{\bf{r}},{\bf{r}}'} \right)} \right] &=& \frac{{\omega ^2 }}{{c^2 }}\left( {\varepsilon  - \frac{{\kappa ^2 }}{\mu }} \right){\cal G}_\omega  \left( {{\bf{r}},{\bf{r}}'} \right) + \delta \left( {{\bf{r}} - {\bf{r}}'} \right)I , \nonumber \\ 
 \nabla  \times \left[ {\frac{1}{\mu }\nabla  \times {\cal G}_\omega  \left( {{\bf{r}},{\bf{r}}'} \right)} \right] &=& \nabla  \times \left[ {\left( {\frac{1}{\mu }\nabla \log \sqrt {\frac{\mu }{\kappa }} } \right) \times {\cal G}_\omega  \left( {{\bf{r}},{\bf{r}}'} \right)} \right],
\end{eqnarray}
where the first of Eqs.(\ref{AbsMirrorG}) is obtained by taking the sum of Eqs.(\ref{Helm}) and (\ref{HelmM}) and the second is deduced by taking the difference of Eqs.(\ref{Helm}) and (\ref{HelmM}), solving the resultant equation for ${\nabla  \times {\cal G}_\omega}\left( {{\bf{r}},{\bf{r}}'} \right)$ and subsequently applying to it the operator $\nabla  \times \left( {\mu ^{ - 1}  \bullet } \right)$. Now Eqs.(\ref{AbsMirrorG}) are manifestly incompatible since the right hand side of the second one, as opposed to the first one, explicitly depends on $\nabla \mu$, $\nabla \kappa$ and the first order spatial derivatives of ${\cal G}_\omega  \left( {{\bf{r}},{\bf{r}}'} \right)$ and in addition it does not contain a delta function. We conclude that ${\cal G}_\omega  \left( {{\bf{r}},{\bf{r}}'} \right)$ and ${\cal G}_\omega ^{\rm M} \left( {{\bf{r}},{\bf{r}}'} \right)$ can not coincide, i.e.
\begin{equation} \label{MAGreen}
{\cal R}{\cal G}_\omega  \left( {{\cal R}{\bf{r}},{\cal R}{\bf{r}}'} \right){\cal R} \ne {\cal G}_\omega  \left( {{\bf{r}},{\bf{r}}'} \right)
\end{equation}
which states the mirror asymmetry of the Green's tensor of a chiral sample. Evidently in an achiral medium where $\kappa = 0$, Eqs.(\ref{Helm}) and (\ref{HelmM}) coincide so that, by uniqueness of solutions, ${\cal G}_\omega ^{\rm M} \left( {{\bf{r}},{\bf{r}}'} \right) = {\cal G}_\omega  \left( {{\bf{r}},{\bf{r}}'} \right)$, i.e. the Green's tensor is mirror symmetric.

Mirror symmetry breaking of the electron quantum properties is a conseguence of the Green's tensor mirror asymmetry. In order to discuss such effects, with reference to Fig.(\ref{FigureS1}), we here consider a chiral medium displaying a reflection symmetry plane $\pi$, the reflection ${\cal R}$ through $\pi$ (so that Eqs.(\ref{MirrorSymmMedium}) are satisfied) and an electron whose velocity $\bf V$ lies on $\pi$ and whose initial wave function is left invariant by ${\cal R}$, i.e.
\begin{equation} \label{phiiMIS}
\phi _i \left( {{\cal R}{\bf{R}}_{\rm t} } \right) = \phi _i \left( {{\bf{R}}_{\rm t} } \right).
\end{equation}
The considered reflection ${\cal R}$, when acting on a vector $\bf F$, does not affect its part parallel to the electron velocity, i.e. ${\cal R}{\bf F}_{\rm v} = {\bf F}_{\rm v} $. In addition, since the transverse part $\bf F_{\rm t}$ lies in the plane orthogonal to $\pi$, it can be decomposed as  ${\bf{F}}_{\rm t}  = {\bf{F}}_\pi   + {\bf{F}}_{\rm n}$ where
\begin{eqnarray} \label{Fpin}
 {\bf{F}}_\pi &=& \frac{1}{2}\left( {{\bf{F}}_{\rm t}  + {\cal R}{\bf{F}}_{\rm t} } \right), \nonumber \\ 
 {\bf{F}}_{\rm n}  &=& \frac{1}{2}\left( {{\bf{F}}_{\rm t}  - {\cal R}{\bf{F}}_{\rm t} } \right) = \frac{1}{2}\left( {{\bf{F}} - {\cal R}{\bf{F}} } \right),
\end{eqnarray}
are its parts parallel ($\pi$) and orthogonal ($\rm n$) to the reflection plane $\pi$ (see Fig.(\ref{FigureS1})). We refer to ${\bf{F}}_{\rm n}$ as the lateral part of the vector $\bf F$. The chosen configuration is geometrically mirror symmetric and the occurrence of any mirror symmetry breaking is accordingly of pure physical origin and due to the medium chirality. Since ${\cal R}{\bf u}_{\rm v} = {\bf u}_{\rm v} $, Eq.(\ref{MAGreen}) readly yields  
\begin{equation}
{\bf{u}}_{\rm v}  \cdot {\cal G}_\omega  \left( {{\cal R}{\bf{r}},{\cal R}{\bf{r}}'} \right){\bf{u}}_{\rm v}  \ne {\bf{u}}_{\rm v}  \cdot {\cal G}_\omega  \left( {{\bf{r}},{\bf{r}}'} \right){\bf{u}} _{\rm v} 
\end{equation}
which combined with the first of Eqs.(\ref{zca}) and Eq.(\ref{DELTAAA}) directly implies that 
\begin{eqnarray} 
\Phi \left( {{\cal R}{\bf{R}}_{\rm t} } \right) &\ne& \Phi \left( {{\bf{R}}_{\rm t} } \right), \nonumber \\
 \Delta \left( {{\cal R}{\bf{R}},{\cal R}{\bf{R}}'} \right) &\ne& \Delta \left( {{\bf{R}},{\bf{R}}'} \right),
\end{eqnarray} 
so that, from Eq.(\ref{gammmm}), we eventually get
\begin{equation} \label{ASMgam}
 \gamma \left( {{\cal R}{\bf{R}},{\cal R}{\bf{R}}'} \right) \ne \gamma \left( {{\bf{R}},{\bf{R}}'} \right).  
\end{equation}
This proves that the mirror symmetry of the electron quantum decoherence is broken by medium chirality. Besides, from the first of Eqs.(\ref{dP3}) we get
\begin{equation}
\left. {\frac{{d \mathscr{P}}}{{d^2 {\bf{P}}_{\rm t} }}} \right|_{{\cal R}{\bf{P}}_{\rm t} }  - \left. {\frac{{d \mathscr{P}}}{{d^2 {\bf{P}}_{\rm t} }}} \right|_{{\bf{P}}_{\rm t} }  = \frac{1}{{\left( {2\pi \hbar } \right)^2 }}\int {d^2 {\bf{R}}_{\rm t} } \int {d^2 {\bf{R}}'_{\rm t} } e^{ - \frac{i}{\hbar }{\bf{P}}_{\rm t}  \cdot \left( {{\bf{R}}_{\rm t}  - {\bf{R}}'_{\rm t} } \right)} \phi _i \left( {{\bf{R}}_{\rm t} } \right)\phi _i^* \left( {{\bf{R}}'_{\rm t} } \right)\left[ {\gamma \left( {{\cal R}{\bf{R}}_{\rm t} ,{\cal R}{\bf{R}}'_{\rm t} } \right) - \gamma \left( {{\bf{R}}_{\rm t} ,{\bf{R}}'_{\rm t} } \right)} \right]
\end{equation}
which, due to the mirror asymmetry of the electron quantum decoherence (see Eq.(\ref{ASMgam})),  remarkably reveals that the transverse momentum distribution is not left invariant by the reflection ${\bf{P}}_{\rm t}  \to {\cal R}{\bf{P}}_{\rm t}$, i.e. it is mirror asymmetric. 

One of the most relevant consequences of this fact is that $\left\langle {{\bf{P}}_{\rm n} } \right\rangle$, the mean value of the lateral momentum, can not be vanishing in spite of the fact that the impinging electron has not such a momentum contribution since $\left\langle {{\bf{P}}_{\rm t} } \right\rangle _i  = \int {d^2 {\bf{R}}_{\rm t} } \phi _i^* \left( {\frac{\hbar }{i}\nabla _{{\bf{R}}_{\rm t} } } \right)\phi _i$ is such that $\left\langle {{\bf{P}}_{\rm t} } \right\rangle _i  = {\cal R}\left\langle {{\bf{P}}_{\rm t} } \right\rangle _i$ by mirror symmetry (see Eq.(\ref{phiiMIS})) and the second of Eqs.(\ref{Fpin}) implies that $\left\langle {{\bf{P}}_{\rm n} } \right\rangle _i = 0$. To discuss this point more in depth we decompose the transverse momentum mean value $\left\langle {{\bf{P}}_{\rm t} } \right\rangle$ in the first of Eqs.(\ref{meanvalues}) and, after some straightforward algebra, we obtain
\begin{eqnarray} \label{avePpin}
 \left\langle {{\bf{P}}_{\pi } } \right\rangle  &=& \int {d^2 {\bf{R}}_{\rm t} } \phi _i^* \left( {\frac{\hbar }{i}\nabla _{{\bf{R}}_{\rm t} } } \right)\phi _i  + \int {d^2 {\bf{R}}_{\rm t} } \left| {\phi _i } \right|^2 \left\{ {\left( {\frac{\hbar }{i}\nabla _{{\bf{R}}_{\rm t} } } \right) \frac{1}{2} \log \left[ {\gamma \left( {{\bf{R}}_{\rm t} ,{\bf{R}}'_{\rm t} } \right)\gamma \left( {{\cal R}{\bf{R}}_{\rm t} ,{\cal R}{\bf{R}}'_{\rm t} } \right)}   \right] } \right\}_{{\bf{R}}'_{\rm t}  = {\bf{R}}_{\rm t} }, \nonumber  \\ 
 \left\langle {{\bf{P}}_{\rm n} } \right\rangle  &=& \int {d^2 {\bf{R}}_{\rm t} } \left| {\phi _i } \right|^2 \left\{ {\left( {\frac{\hbar }{i}\nabla _{{\bf{R}}_{\rm t} } } \right) \frac{1}{2} \log \left[ {  {\frac{{\gamma \left( {{\bf{R}}_{\rm t} ,{\bf{R}}'_{\rm t} } \right)}}{{\gamma \left( {{\cal R}{\bf{R}}_{\rm t} ,{\cal R}{\bf{R}}'_{\rm t} } \right)}}} } \right]} \right\}_{{\bf{R}}'_{\rm t}  = {\bf{R}}_{\rm t} }.
\end{eqnarray}
The first of these equations states that the contribution to $\left\langle {{\bf{P}}_{\pi } } \right\rangle$ that the electron gets as a result of its interaction with the medium (the second term in the right hand side) is generally non-vanishing and it is independent on the medium chirality since it survives in the achiral case where $\gamma \left( {{\cal R}{\bf{R}}_{\rm t} ,{\cal R}{\bf{R}}'_{\rm t} } \right) = \gamma \left( {{\bf{R}}_{\rm t} ,{\bf{R}}'_{\rm t} } \right)$. On the other hand, the second of Eqs.(\ref{avePpin}) dramatically states that $\left\langle {{\bf{P}}_{\rm n} } \right\rangle$ is here solely provided by the electron-medium interaction and that it does not vanish only if the quantum electron decoherence is mirror asymmetric $\gamma \left( {{\cal R}{\bf{R}}_{\rm t} ,{\cal R}{\bf{R}}'_{\rm t} } \right) \ne \gamma \left( {{\bf{R}}_{\rm t} ,{\bf{R}}'_{\rm t} } \right)$, that is to say if the medium is chiral.

The non-vanishing lateral momentum $\left\langle {{\bf{P}}_{\rm n} } \right\rangle$ transferred to the electron is related to a mechanical interaction that the chiral medium is able to exert on the electron along the direction normal to the reflection symmetry plane $\pi$. In spite of this lateral interaction, in our description the mean value of the electron transverse position $\left\langle {{\bf{R}}_{\rm t} } \right\rangle  = {\rm Tr}  ( {\hat \rho _e {\bf{\hat R}}_{\rm t} } )$ does not change upon scattering since, from Eq.(\ref{rhoe}), $\rho _e \left( {{\bf{R}},{\bf{R}}} \right) = \left| {\psi _i \left( {\bf{R}} \right)} \right|^2$ and hence
\begin{equation}
\left\langle {{\bf{R}}_{\rm t} } \right\rangle  = \int {d^3 {\bf{R}}} \left| {\phi _i \left( {\bf{R}} \right)} \right|^2 {\bf{R}}_{\rm t} 
\end{equation}
which is the mean value of the transverse position of the initial electron wavefunction. This is due to the paraxial aproximation adopted to describe the electron where the electron Hamiltonian is ${\bf{V}} \cdot {\bf{\hat P}} = V\hat P_{\rm v}$ (see Eq.(\ref{H2-1})) which evidently commutes with the transverse position operator ${{\bf{\hat R}}_{\rm t} }$. Accordingly ${{\bf{\hat R}}_{\rm t} }$ commutes with the full Hamiltonian $\hat H$ of Eq.(\ref{H2-1}) and its mean value can not vary due the Ehrenfest theorem
\begin{equation}
\frac{{d\left\langle {{\bf{R}}_{\rm t} } \right\rangle }}{{dt}} = \frac{1}{{i\hbar }}\left\langle {\Psi \left( t \right)} \right|\left[ {{\bf{\hat R}}_{\rm t} ,\hat H} \right]\left| {\Psi \left( t \right)} \right\rangle. 
\end{equation}

\label{SE2}

\subsection{Weak coupling regime} 
The reduced density matrix in Eq.(\ref{rhoe}) and the transverse momentum distribution in Eq.(\ref{dP3}) have been rigorously evaluated by using $S$-operator of Eq.(\ref{Sop2}) which in turn has been obtained by solving the time evolution equation in Eq.(\ref{Evol1}) with no approximation. As a consequence our treatment is non-perturbative and it enables the investigation of the electron-chiral medium interaction beyond the weak coupling regime. For example, this can occur by letting the electron to fly near a large chiral sample providing a long interaction length (see below). This possibilty is very important for our purposes since it could ehance the mirror symmetry breaking effects discussed in Sec.S2.5. In fact such asymmetry effects are a physical consequence of medium chirality which is generally very weak (the Pasteur parameter in Eq.(\ref{consti}) is typically $|\kappa| \sim 10^{-4}$) and accordingly the asymmetry is expected to be very tiny in the weak coupling regime. 

To discuss the weak coupling regime, we start by examining the full electron-radiation state of Eq.(\ref{Psi1}) which turns out to be the superposition of an infinite number of states, the nth of which having precisely $n$ elementary polaritonic excitations and being an eigenstate of eigenvalue $E_i$ of the free electron-radiation Hamiltonian ${\hat H}_0$ due to the relation
\begin{equation}
\left[ {\hat H_0 ,\int {d\xi } \;T ( {{\bf{\hat R}},\xi }  )\hat f^\dag  \left( \xi  \right)} \right] = 0,
\end{equation}
which can easily be obtained from the first of Eqs.(\ref{H3}) and the third of Eqs.(\ref{T}) with the help of the basic electron position-momentum commutation relations and the polaritonic commutation relations of Eqs.(\ref{commf}). As a consequence the term with $n=0$ is basically a renormalization of the bare electron initial state whereas the nth term describes a process where the electron loses energy due to the generation of $n$ polaritonic excitations. The weak coupling regime occurs when the only relevant process is the generation of a single polaritonic excitation and accordingly all the terms with $n >1$ can be neglected, a situation described by the standard first order perturbation theory. Now Eq.(\ref{rhoe0}) reveals that the reduced density matrix gets a contribution from each of the n-polaritonic excitation process and that higher order terms can be neglected if
\begin{equation}
\left| {\Delta \left( {{\bf{R}},{\bf{R}}'} \right)} \right| \ll 1.
\end{equation}
which accordingly is the condition characterizing the weak coupling regime. This condition is also accompanied by the fact that $\left| {\Phi \left( {\bf{R}} \right)} \right| \ll 1$ and hence the first order expansion of the reduced density matrix reads
\begin{eqnarray}
 \rho _e \left( {{\bf{R}},{\bf{R}}'} \right) &\simeq& \psi _i \left( {\bf{R}} \right)\psi _i^* \left( {{\bf{R}}'} \right)\left[ {1 + \Delta \left( {{\bf{R}},{\bf{R}}'} \right)} \right] = \nonumber \\ 
& =& \psi _i \left( {\bf{R}} \right)\psi _i^* \left( {{\bf{R}}'} \right)\left\{ {1 + \frac{{4\alpha }}{c}\int\limits_0^\infty  {d\omega } e^{ - i\frac{\omega }{V}\left( {R_{\rm v}  - R'_{\rm v} } \right)} \int\limits_{ - \infty }^{ + \infty } {dq} \int\limits_{ - \infty }^{ + \infty } {dq'} e^{i\frac{\omega }{V}\left( {q - q'} \right)}  {\mathop{\rm Im}\nolimits} \left[ {\bf{u}}_{\rm v}  \cdot {{\cal G}_\omega  \left( {{\bf{R}}_{\rm t}  + q{\bf{u}}_{\rm v} ,{\bf{R}}'_{\rm t}  + q'{\bf{u}}_{\rm v} } \right)} {\bf{u}}_{\rm v} \right] } \right\}, \nonumber \\ 
\end{eqnarray}
which formally coincides with the results of Ref.\cite{Mechel} where the authors consider the interaction of fast electrons with non magnetoelectric media. Analagously, the first order expansion of the electron energy distribution in the second of Eqs.(\ref{dP3}) is
\begin{equation}
\frac{{d \mathscr{P}}}{{dE}} \simeq \delta \left( {E - E_i } \right) + \Gamma \left( {\frac{{E_i  - E}}{\hbar }} \right)
\end{equation}
where
\begin{equation}
\Gamma \left( \omega  \right) = \frac{{4\alpha }}{{c\hbar }}\int {d^2 {\bf{R}}_{\rm t} } \left| {\phi _i \left( {{\bf{R}}_{\rm t} } \right)} \right|^2 \int\limits_{ - \infty }^{ + \infty } {dq} \;\int\limits_{ - \infty }^{ + \infty } {dq'} e^{i\frac{\omega }{V}\left( {q - q'} \right)}  {\mathop{\rm Im}\nolimits} \left[ {\bf{u}}_{\rm v}  \cdot {{\cal G}_\omega  \left( {{\bf{R}}_{\rm t}  + q{\bf{u}}_{\rm v} ,{\bf{R}}_{\rm t}  + q'{\bf{u}}_{\rm v} } \right)} {\bf{u}}_{\rm v} \right] 
\end{equation}
is the well-knonw expression of the electron energy loss probability pertaining non magnetoelectric media \cite{deAba}.

\section{Aloof electron traveling parallel to a chiral film}
In order to specialize the above general results we hereafter focus on an aloof electron traveling parallel to a chiral film. With reference to left side of Fig.\ref{FigureS2}, the homogenous chiral film of thickness $d$, permittivity $\varepsilon$, permeability $\mu \simeq 1$ and Pasteur parameter $\kappa$ is in vacuum ($\varepsilon_1 = 1$) and it lies onto the substrate of permittivity $\varepsilon_2$. The medium (substrate + chiral film + vacuum) has the mirror symmetry plane $\pi = zx$ and  the reflection through such plane is provided by the tensor
\begin{equation} \label{Ryyyyy}
{\cal R} = {\bf{e}}_x {\bf{e}}_x  - {\bf{e}}_y {\bf{e}}_y  + {\bf{e}}_z {\bf{e}}_z.
\end{equation}
The electron has initial mirror symmetric wavefunction  $\psi _i \left( {{\cal R}{\bf{R}}} \right) = \psi _i \left( {\bf{R}} \right)$ and its velocity is along the $x$ axis, so that ${\bf{u}}_{\rm v}  = {\bf{e}}_x$. Such configuration is of the kind used in Sec.2.5. to discuss the mirror symmetry breaking effects and the decomposition of a vector $\bf F$ described in Fig.(\ref{FigureS1}) here becomes
\begin{eqnarray}
 {\bf{F}}_{\rm v}  = F_x {\bf{e}}_x, \nonumber  \\ 
 {\bf{F}}_\pi   = F_z {\bf{e}}_z, \nonumber  \\ 
 {\bf{F}}_{\rm n}  = F_y {\bf{e}}_y,
\end{eqnarray}
so that asymmetric effects show up along the $y$-axis which is the lateral direction. In the chosen aloof configuration the initial electron wavefunction $\phi _i \left( Y,Z \right) $ is non-vanishing only in vacuum ($Z<0$) so that, from Eq.(\ref{rhoe}), the electron reduced density matrix does not vanish only in vacuum as well. Accordingly we will evaluate the Green's tensor component ${\bf{e}}_x  \cdot {\cal G}_\omega  \left( {{\bf{r}},{\bf{r}}'} \right){\bf{e}}_x$ and the fundamental quantities $\Phi ( {\bf{R}} )$ and ${\Delta ( {{\bf{R}},{\bf{R}}'} )}$ only for $z<0$, $z'<0$, $Z<0$ and $Z'<0$. For the sake of clarity, in the right side of Fig.\ref{FigureS2} we have summarized the quantities used in the below analysis.

\subsection{Green's tensor component along the electron velocity}
\begin{figure}
\begin{minipage}{.55\columnwidth}
\centering
\includegraphics[width = 0.95\linewidth]{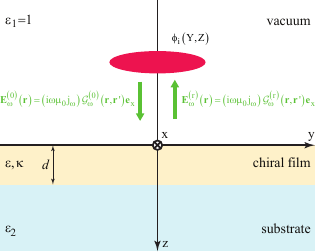}
\end{minipage}\hfill
\begin{minipage}{0.45\columnwidth}
\begin{eqnarray}
   {{\bf{r}}_\parallel   = x{\bf{e}}_x  + y{\bf{e}}_y }      \: :
   && \quad {\rm position \: vector}        \nonumber \\
   && \quad {\rm parallel \: part}        \nonumber \\
   {{\bf{r}}'_\parallel   = x'{\bf{e}}_x  + y'{\bf{e}}_y }      \: :
   && \quad {\rm source \: vector}        \nonumber \\
   && \quad {\rm parallel \: part}        \nonumber \\
   {{\bf{k}}_\parallel   = k_x {\bf{e}}_x  + k_y {\bf{e}}_y } \: :
   && \quad {\rm parallel \: wavevector}                \nonumber \\
   k_\omega = \omega/c \: :
   && \quad {\rm vacuum \: vavenumber}  \nonumber \\ 
   k_{1z}  = \sqrt {k_\omega^2 \varepsilon _1  - k_\parallel ^2 } \: :
   && \quad {\rm vacuum}  \nonumber \\ 
   && \quad {\rm longitudinal \: wavenumber}  \nonumber \\ 
   k_{2z}  = \sqrt {k_\omega^2 \varepsilon _2  - k_\parallel ^2 } \: :
   && \quad {\rm substrate}  \nonumber \\
   && \quad {\rm longitudinal \: wavenumber}  \nonumber \\
   {\bf{u}}_{\rm S}  = {\bf{e}}_z  \times {{{\bf{k}}_\parallel  }}/{{k_\parallel  }} \: :
   && \quad {\rm S \: unit \: vector}  \nonumber \\
   {\bf{u}}_{\rm P}  = {{\bf{k}}_\parallel  } /{{k_\parallel  }}  \: :
   && \quad {\rm P \: unit \: vector}  \nonumber \\
   n = {\sqrt {\varepsilon \kappa ^2 } }/{\kappa } \: :
   && \quad {\rm chiral \: refractive \: index}  \nonumber \\
   K_{z \pm }  = \sqrt {k_\omega^2 \left( {n \pm \kappa } \right)^2  - k_\parallel ^2 } \: :
   && \quad {\rm chiral \: film}  \nonumber \\
   && \quad {\rm longitudinal \:  wavenumbers}  \nonumber 
\end{eqnarray}
\end{minipage}
\caption{{\bf Left:} Geometry of the interaction between 
the aloof electron traveling in vacuum parallel to the $x$ axis nearby the chiral film deposited onto the substrate. The red spot displays the initial electron wavefunction transverse profile $\phi_i(Y,Z)$ whereas the green arrows sketch the vacuum and reflected fields, ${\bf{E}}_\omega ^{\left( 0 \right)}$ and ${\bf{E}}_\omega ^{\left( r \right)}$, used to evaluate the Green's tensor component along the electron velocity. {\bf Right:} List of quantities used in the evaluation.}
\label{FigureS2}
\end{figure}

In order to evaluate the Green's tensor component ${\bf{e}}_x  \cdot {\cal G}_\omega  \left( {{\bf{r}},{\bf{r}}'} \right){\bf{e}}_x$ in the vacuum half-space we start noting that an elementary current density ${\bf{J}}_\omega  \left( {\bf{r}} \right) = J_\omega  \delta \left( {{\bf{r}} - {\bf{r}}'} \right){\bf{e}}_x$ located in vacuum ($z'<0$) produces in $z<0$ the field 
\begin{equation} \label{EEE}
{\bf{E}}_\omega  \left( {\bf{r}} \right) = \left( {i\omega \mu _0 J_\omega  } \right)\;{\cal G}_\omega  \left( {{\bf{r}},{\bf{r}}'} \right){\bf{e}}_x  = {\bf{E}}_\omega ^{\left( 0 \right)} \left( {\bf{r}} \right) + {\bf{E}}_\omega ^{\left( r \right)} \left( {\bf{r}} \right)
\end{equation}
where ${\bf{E}}_\omega ^{\left( 0 \right)}$ is the free-space field produced by the elementary current 
\begin{equation} \label{Ei00}
{\bf{E}}_\omega ^{\left( 0 \right)}  = \left( {i\omega \mu _0 J_\omega  } \right)\left( {1 + \frac{{\nabla \nabla \cdot }}{{k_\omega^2 \varepsilon _1 }}} \right)\frac{{e^{ik_\omega \sqrt {\varepsilon _1 } \left| {{\bf{r}} - {\bf{r}}'} \right|} }}{{4\pi \left| {{\bf{r}} - {\bf{r}}'} \right|}}{\bf{e}}_x
\end{equation}
and ${\bf{E}}_\omega ^{\left( r \right)}$ is the field reflected by the chiral film interface (see left side of Fig.\ref{FigureS2}). To evaluate the reflected field it is convenient to cast the free-space field as
\begin{equation} \label{Ei0}
{\bf{E}}_\omega ^{\left( 0 \right)}  = \left( {i\omega \mu _0 J_\omega  } \right) \frac{i}{{8\pi ^2 }} \int {d^2 {\bf{k}}_\parallel  e^{i{\bf{k}}_\parallel   \cdot \left( {{\bf{r}}_\parallel   - {\bf{r}}'_\parallel  } \right)}  \frac{e^{ik_{1z} \left| {z - z'} \right|}}{{ k_{1z} }}} \left\{ {\left( { - \frac{{k_y }}{{k_\parallel  }}} \right){\bf{u}}_{\rm S}  + \left( {\frac{{k_{1z}^2 k_x }}{{k_\parallel  k_\omega^2 \varepsilon _1 }}} \right)\left[ {{\bf{u}}_{\rm P}  - \frac{{k_\parallel  }}{{k_{1z} }}{\mathop{\rm sgn}} \left( {z - z'} \right){\bf{e}}_z } \right]} \right\}
\end{equation}
where we have used the Weyl representation of the spherical wave
\begin{equation}
\frac{{e^{ik_\omega \sqrt {\varepsilon _1 } \left| {{\bf{r}} - {\bf{r}}'} \right|} }}{{\left| {{\bf{r}} - {\bf{r}}'} \right|}} = \frac{i}{{2\pi }}\int {d^2 {\bf{k}}_\parallel  e^{i{\bf{k}}_\parallel   \cdot \left( {{\bf{r}}_\parallel   - {\bf{r}}'_\parallel  } \right)} } \frac{{e^{ik_{1z} \left| {z - z'} \right|} }}{{k_{1z} }}.
\end{equation} 
Now in Ref.\cite{Ciatt1} it has been shown that an incident field 
\begin{equation} \label{Ei}
{\bf{E}}_\omega ^{\left( i \right)}  = \int {d^2 {\bf{k}}_\parallel  } e^{i{\bf{k}}_\parallel   \cdot {\bf{r}}_\parallel  } e^{ik_{1z} z} \left[ {U_{\rm S}^{\left( i \right)} {\bf{u}}_{\rm S}  + U_{\rm P}^{\left( i \right)} \left( {{\bf{u}}_{\rm P}  - \frac{{k_\parallel  }}{{k_{1z} }}{\bf{e}}_z } \right)} \right]
\end{equation}
impinging onto the chiral film/substrate composite produces in $z<0$ the reflected field
\begin{equation} \label{Er}
{\bf{E}}_\omega ^{\left( r \right)}  = \int {d^2 {\bf{k}}_\parallel  } e^{i{\bf{k}}_\parallel   \cdot {\bf{r}}_\parallel  } e^{ - ik_{1z} z} \left[ {\left( {R_{\rm SS} U_{\rm S}^{\left( i \right)}  + nR_{\rm SP} U_{\rm P}^{\left( i \right)} } \right){\bf{u}}_{\rm S}  + \left( {nR_{\rm PS} U_{\rm S}^{\left( i \right)}  + R_{\rm PP} U_{\rm P}^{\left( i \right)} } \right)\left( {{\bf{u}}_{\rm P}  + \frac{{k_\parallel  }}{{k_{1z} }}{\bf{e}}_z } \right)} \right]
\end{equation}
in which the reflection matrix is given by
\begin{equation} \label{RefMat}
\left( {\begin{array}{@{\mkern0mu} c c @{\mkern0mu}} 
   {R_{\rm SS} } & {nR_{\rm SP} }  \\
   {nR_{\rm PS} } & {R_{\rm PP} }  \\
\end{array}} \right) = \left( {\begin{array}{@{\mkern0mu} c c @{\mkern0mu}}
   1 & 0  \\
   0 & { - 1}  \\
\end{array}} \right)\left( {M^{\left( 1 \right)}  - M^{\left( 2 \right)} } \right)\left( {M^{\left( 1 \right)}  + M^{\left( 2 \right)} } \right)^{ - 1} 
\end{equation}
where

\begin{eqnarray} \label{MatrM1M2}
M^{\left( 1 \right)}  &=& \left( {\begin{array}{@{\mkern0mu} c c @{\mkern0mu}}
   {\left( {C_ +   + \frac{{k_{2z} }}{{ik_\omega }}S_ + ^{\left( 1 \right)} } \right)} & {n\left( {\frac{{k_\omega \varepsilon _2 }}{{ik_{2z} \varepsilon }}C_ -   - S_ - ^{\left( 1 \right)} } \right)}  \\
   {n\frac{{ik_{1z} }}{{k_\omega \varepsilon _1 }}\left( {C_ -   + \frac{{k_{2z} }}{{ik_\omega }}S_ - ^{\left( 1 \right)} } \right)} & {\varepsilon \frac{{ik_{1z} }}{{k_\omega \varepsilon _1 }}\left( {\frac{{k_\omega \varepsilon _2 }}{{ik_{2z} \varepsilon }}C_ +   - S_ + ^{\left( 1 \right)} } \right)}  \\
\end{array}} \right), \nonumber \\
M^{\left( 2 \right)}  &=& \left( {\begin{array}{@{\mkern0mu} c c @{\mkern0mu}}
   {\frac{{k_\omega }}{{ik_{1z} }}\left( {\frac{{ik_{2z} }}{{k_\omega }}C_ +   + S_ + ^{\left( 2 \right)} } \right)} & {n\frac{{k_\omega }}{{ik_{1z} }}\left( {C_ -   + \frac{{k_\omega \varepsilon _2 }}{{ik_{2z} \varepsilon }}S_ - ^{\left( 2 \right)} } \right)}  \\
   {n\frac{1}{\varepsilon }\left( {\frac{{ik_{2z} }}{{k_\omega }}C_ -   + S_ - ^{\left( 2 \right)} } \right)} & {\left( {C_ +   + \frac{{k_\omega \varepsilon _2 }}{{ik_{2z} \varepsilon }}S_ + ^{\left( 2 \right)} } \right)}  \\
\end{array}} \right),
\end{eqnarray}
and
\begin{eqnarray}
 C_ \pm   &=& \frac{1}{2}\left[ {\cos \left( {K_{z + } d} \right) \pm \cos \left( {K_{z - } d} \right)} \right], \nonumber \\ 
 S_ \pm ^{\left( 1 \right)}  &=& \frac{1}{2}\left[ {\frac{{k_\omega \left( {n + \kappa } \right)}}{{K_{z + } n}}\sin \left( {K_{z + } d} \right) \pm \frac{{k_\omega \left( {n - \kappa } \right)}}{{K_{z - } n}}\sin \left( {K_{z - } d} \right)} \right], \nonumber  \\ 
 S_ \pm ^{\left( 2 \right)}  &=& \frac{1}{2}\left[ {\frac{{K_{z + } n}}{{k_\omega \left( {n + \kappa } \right)}}\sin \left( {K_{z + } d} \right) \pm \frac{{K_{z - } n}}{{k_\omega \left( {n - \kappa } \right)}}\sin \left( {K_{z - } d} \right)} \right].
\end{eqnarray}
Is is worth noting that the reflection coefficients $R_{\rm IJ}$ defined in Eq.(\ref{RefMat}) are left invariant by chirality reversal $\kappa \rightarrow -\kappa$ as can easily be proved with help of the matrices $M^{(j)}$ of Eqs.(\ref{MatrM1M2}). The only term affected by chirality reversal is the chiral refractive index $n = \sqrt{\varepsilon \kappa^2}/\kappa$ whose sign is switched by the reversal. 

Since along the stripe $z' < z < 0$ the free-space field in Eq.(\ref{Ei0}) has the same structure of the field in Eq.(\ref{Ei}) with $\rm S$ and $\rm P$ amplitudes
\begin{eqnarray}
 U_{\rm S}^{\left( i \right)}  &=& \left( {i\omega \mu _0 J_\omega  } \right)\frac{i}{{8\pi ^2 }}e^{ - i{\bf{k}}_\parallel   \cdot {\bf{r}}'_\parallel  } \frac{{e^{ - ik_{1z} z'} }}{{k_{1z} }}\left( { - \frac{{k_y }}{{k_\parallel  }}} \right), \nonumber \\ 
 U_{\rm P}^{\left( i \right)}  &=& \left( {i\omega \mu _0 J_\omega  } \right)\frac{i}{{8\pi ^2 }}e^{ - i{\bf{k}}_\parallel   \cdot {\bf{r}}'_\parallel  } \;\frac{{e^{ - ik_{1z} z'} }}{{k_{1z} }}\left( {\frac{{k_{1z}^2 k_x }}{{k_\parallel  k_\omega^2 \varepsilon _1 }}} \right), 
\end{eqnarray}
the reflected field is readily given by Eq.(\ref{Er}) and it is 
\begin{eqnarray} \label{Er1}
 {\bf{E}}_\omega ^{\left( r \right)}  &=& \left( {i\omega \mu _0 J_\omega  } \right) \frac{i}{{8\pi ^2 }}   \int {d^2 {\bf{k}}_\parallel  } e^{i{\bf{k}}_\parallel   \cdot \left( {{\bf{r}}_\parallel   - {\bf{r}}'_\parallel  } \right)}   \frac{e^{ - ik_{1z} \left( {z + z'} \right)}}{{ k_{1z} }}  \left[ {\left( { - R_{\rm SS} \frac{{k_y }}{{k_\parallel  }} + nR_{\rm SP} \frac{{k_{1z}^2 k_x }}{{k_\parallel  k_\omega^2 \varepsilon _1 }}} \right){\bf{u}}_{\rm S} } + \right. \nonumber \\ 
&+&  \left.  {\left( { - nR_{\rm PS} \frac{{k_y }}{{k_\parallel  }} + R_{\rm PP} \frac{{k_{1z}^2 k_x }}{{k_\parallel  k_\omega^2 \varepsilon _1 }}} \right)\left( {{\bf{u}}_{\rm P}  + \frac{{k_\parallel  }}{{k_{1z} }}{\bf{e}}_z } \right)} \right].
\end{eqnarray}
Projecting the field in Eq.(\ref{EEE}) onto the $x$ axis and using Eqs.(\ref{Ei0}) and (\ref{Er1}) we get the xx component ${\cal G}_{\omega xx}  = {\bf{e}}_x  \cdot {\cal G}_\omega  {\bf{e}}_x$ of the Green's tensor 
\begin{equation} \label{GGxx}
{\cal G}_{\omega xx} \left( {{\bf{r}},{\bf{r}}'} \right) = {\cal G}_{\omega xx}^{\left( 0 \right)} \left( {{\bf{r}},{\bf{r}}'} \right) + {\cal G}_{\omega xx}^{\left( r \right)} \left( {{\bf{r}},{\bf{r}}'} \right)
\end{equation}
where 
\begin{align} \label{Gxx}
{\cal G}_{\omega xx}^{\left( 0 \right)} \left( {{\bf{r}},{\bf{r}}'} \right)  &= \frac{i}{{8\pi ^2 }}\int {d^2 {\bf{k}}_\parallel  e^{i{\bf{k}}_\parallel   \cdot \left( {{\bf{r}}_\parallel   - {\bf{r}}'_\parallel  } \right)} \frac{{e^{ik_{1z} \left| {z - z'} \right|} }}{{k_{1z} }}} \left( {1 - \frac{{k_x^2 }}{{k_\omega^2 \varepsilon _1 }}} \right), \nonumber \\ 
{\cal G}_{\omega xx}^{\left( r \right)} \left( {{\bf{r}},{\bf{r}}'} \right)  &= \frac{i}{{8\pi ^2 }}\int {d^2 {\bf{k}}_\parallel  } e^{i{\bf{k}}_\parallel   \cdot \left( {{\bf{r}}_\parallel   - {\bf{r}}'_\parallel  } \right)} \frac{{e^{ - ik_{1z} \left( {z + z'} \right)} }}{{k_{1z} }}\left[ {\left( {R_{\rm SS} \frac{{k_y^2 }}{{k_\parallel ^2 }} + R_{\rm PP} \frac{{k_{1z}^2 }}{{k_\omega^2 \varepsilon _1 }}\frac{{k_x^2 }}{{k_\parallel ^2 }}} \right) - n\left( {R_{\rm SP} \frac{{k_{1z}^2 }}{{k_\omega^2 \varepsilon _1 }} + R_{\rm PS} } \right)\frac{{k_x k_y }}{{k_\parallel ^2 }}} \right] \nonumber \\ 
\end{align}
are the $xx$ components of the free-space and reflected parts of the Green's tensor, ${\cal G}_\omega ^{\left( 0 \right)} \left( {{\bf{r}},{\bf{r}}'} \right)$ and ${\cal G}_\omega ^{\left( r \right)} \left( {{\bf{r}},{\bf{r}}'} \right)$.

\subsection{Mirror asymmetry of the Green's tensor as due to mirror optical activity}
Green's tensor mirror asymmetry of a geometrically mirror symmetric chiral medium (see Eq.(\ref{MAGreen})) is a very general property which is independent on the medium shape or inhomegeneity. In the specific configuration we are examining here (left side of Fig.\ref{FigureS2}) such mirror asymmetry can be related to mirror optical activity (MOA), a chiroptical phenomenon investigated in Refs.\cite{Ciatt1, Ciatt2} where the field reflected by a homogeneous chiral film ${\bf{E}}_\omega ^{\left( r \right)}$ has always indefinite parity (asymmetry) even if the incident field ${\bf{E}}_\omega ^{\left( i \right)}$ is mirror symmetric. To show such connection and gain further insight about Green's tensor asymmetry, we start by briefly summarizing MOA description. We say that a field ${\bf{E}}_\omega  \left( {\bf{r}} \right)$ is a mirror symmetric field (MSF) if it coincides with its mirror image ${\bf{E}}^{\rm M}_\omega  \left( {\bf{r}} \right) = {\cal R}{\bf{E}}_\omega  \left( {{\cal R}{\bf{r}}} \right)$, i.e. ${\bf{E}}_\omega  \left( {\bf{r}} \right) = {\bf{E}}^{\rm M}_\omega  \left( {\bf{r}} \right)$. For the reflection through the $xz$ we are considering (see Eq.(\ref{Ryyyyy})) the $\rm S$ and $\rm P$ unit vectors (see right side of Fig.\ref{FigureS2}) evidently satisfy the momentum space symmetry relations
\begin{eqnarray}
 {\cal R}{\bf{u}}_{\rm S} \left( {{\cal R}{\bf{k}}_\parallel  } \right) &=&  - {\bf{u}}_{\rm S} \left( {{\bf{k}}_\parallel  } \right), \nonumber \\ 
 {\cal R}{\bf{u}}_{\rm P} \left( {{\cal R}{\bf{k}}_\parallel  } \right) &=& {\bf{u}}_{\rm P} \left( {{\bf{k}}_\parallel  } \right),
\end{eqnarray}
so that it is straightforward deducing from Eq.(\ref{Ei}) that ${\bf{E}}_\omega ^{\left( i \right)}$ is a MSF if and only if it has antisymmetric and symmetric $\rm S$ and $\rm P$ amplitudes respectively 
\begin{eqnarray}
U_{\rm S}^{\left( i \right)} \left( {{\cal R}{\bf{k}}_\parallel  } \right) &=& - U_{\rm S}^{\left( i \right)} \left( {{\bf{k}}_\parallel  } \right), \nonumber \\
U_{\rm P}^{\left( i \right)} \left( {{\cal R}{\bf{k}}_\parallel  } \right) &=& U_{\rm P}^{\left( i \right)} \left( {{\bf{k}}_\parallel  } \right).
\end{eqnarray}
Such momentum space characterization of the incident MSF enables to prove that the reflected field of Eq.(\ref{Er}) is such that
\begin{equation} \label{asymmErrr}
{\bf{E}}_\omega ^{\left( r \right)} - {\bf{E}}_\omega ^{\left( r \right) {\rm M}}  =   2n\int {d^2 {\bf{k}}_\parallel  } e^{i{\bf{k}}_\parallel   \cdot {\bf{r}}_\parallel  } e^{ - ik_{1z} z} \left[ {R_{\rm SP} U_{\rm P}^{\left( i \right)} {\bf{u}}_{\rm S}  + R_{\rm PS} U_{\rm S}^{\left( i \right)} \left( {{\bf{u}}_{\rm P}  + \frac{{k_\parallel  }}{{k_{1z} }}{\bf{e}}_z } \right)} \right]
\end{equation}
where use has been made of the fact that the reflection coefficients $R_{\rm IJ} ( {k_\parallel ^2 } )$ in Eq.(\ref{RefMat}) are rotationally invariant around the $z$-axis. Now the right hand side of Eq.(\ref{asymmErrr}) does not vanish due to the mixing reflection coefficients $R_{\rm SP}$ and $R_{\rm PS}$ which result from the S-P coupling mediaded by chirality and thus vanishing in a achiral medium (see Ref.\cite{Ciatt1} for further details). Therefore, the reflected field ${\bf{E}}_\omega ^{\left( r \right)}$ is not a MSF and MOA is described by the $R_{\rm SP}$ and $R_{\rm PS}$ reflection coefficients.

To show the impact of MOA on the Green's tensor $xx$ component evaluated in the above section, it is sufficient to observe that the free-space field ${\bf{E}}_\omega ^{\left( 0 \right)}$ of Eq.(\ref{Ei00}) is produced by an elementary current parallel to the $x$ axis and located at $\bf r'$ so that it is a MSF with respect the reflection through the plane orthogonal to the $y$ axis and containing $\bf r'$, namely
\begin{equation}
  ({{\bf{r}}_\parallel   - {\bf{r}}'_\parallel  }  )^{\rm M}  = {\cal R} ( {{\bf{r}}_\parallel   - {\bf{r}}'_\parallel  }  ).
\end{equation}
Accordingly Eq.(\ref{Ei0}) shows that ${\bf{E}}_\omega ^{\left( 0 \right)}$ depends on $({{\bf{r}}_\parallel   - {\bf{r}}'_\parallel  })$ and that its $\rm S$ and $\rm P$ amplitudes are antisymmetric and symmetric, respectively, this explaining the symmetry of the free-space part of the Green's tensor $xx$ component  
\begin{equation}
{\cal G}_{\omega xx}^{\left( 0 \right)} \left( {{\cal R}{\bf{r}},{\cal R}{\bf{r}}'} \right) = {\cal G}_{\omega xx}^{\left( 0 \right)} \left( {{\bf{r}},{\bf{r}}'} \right)
\end{equation}
which is evident from the first of Eqs.(\ref{Gxx}). Due to MOA the reflected field ${\bf{E}}_\omega ^{\left( r \right)}$ of Eq.(\ref{Er1}) is not a MSF and this entails the asymmetry of the reflected part of the Green's tensor $xx$ component. As a matter of fact from the second of Eqs.(\ref{Gxx}) we get
\begin{equation}
 {\cal G}_{\omega xx}^{\left( r \right)} \left( {{\cal R}{\bf{r}},{\cal R}{\bf{r}}'} \right) - {\cal G}_{\omega xx}^{\left( r \right)} \left( {{\bf{r}},{\bf{r}}'} \right)  
 = 2n\frac{i}{{8\pi ^2 }}\int {d^2 {\bf{k}}_\parallel  } e^{i{\bf{k}}_\parallel   \cdot \left( {{\bf{r}}_\parallel   - {\bf{r}}'_\parallel  } \right)} \frac{{e^{ - ik_{1z} \left( {z + z'} \right)} }}{{k_{1z} }}\left( {R_{\rm SP} \frac{{k_{1z}^2 }}{{k_\omega^2 \varepsilon _1 }} + R_{\rm PS} } \right)\frac{{k_x k_y }}{{k_\parallel ^2 }} 
\end{equation}
clearly revealing that the general mirror asymmetry of the Green's tensor stated in Eq.(\ref{MAGreen}) is here due to MOA since its right hand side only contains the mixing reflection coefficients $R_{\rm SP}$ and $R_{\rm PS}$.

\subsection{Evaluation of the electron reduced density matrix}
We evaluate here the phase factor $\Phi \left( {{\bf{R}} } \right)$ and the fundamental quantity $\Delta \left( {{\bf{R}},{\bf{R}}'} \right)$ for $Z<0$ and $Z'<0$. Since the electron velocity is parallel to the $x$ axis,  the $\rm t$ and $\rm v$ parts of $\bf{R}$ and $\bf{R}'$ are (see Eq.(\ref{decomp})) 
\begin{equation}
\renewcommand{\arraystretch}{1.5}
\begin{array}{rlcrl}
   {\bf{R}}_{\rm t}  &=& Y{\bf{e}}_y   + Z{\bf{e}}_z,  & & R_{\rm v}   = X, \\
  {\bf{R}}'_{\rm t}  &=& Y'{\bf{e}}_y  + Z'{\bf{e}}_z, & & R'_{\rm v}   = X',  \\
\end{array}
\end{equation}
and accordingly, from the first of Eqs.(\ref{zca}) and Eq.(\ref{DELTAAA}) we have
\begin{eqnarray} \label{zeDe0}
 \Phi \left( {{\bf{R}}_{\rm t} } \right) &=& \frac{{4\alpha }}{c}{\mathop{\rm Im}\nolimits} \int\limits_0^{ + \infty } {d\omega } \int\limits_{ - \infty }^{ + \infty } {dq} \int\limits_{ - \infty }^q {dq'\,} e^{i\frac{\omega }{V}\left( {q - q'} \right)} {\mathop{\rm Im}\nolimits} \left[ {{\cal G}_{\omega xx} \left( {{\bf{r}},{\bf{r}}'} \right)} \right]_{\scriptstyle {\bf{r}} = q{\bf{e}}_x  + Y{\bf{e}}_y  + Z{\bf{e}}_z  \hfill \atop 
  \scriptstyle {\bf{r}}' = q'{\bf{e}}_x  + Y{\bf{e}}_y  + Z{\bf{e}}_z  \hfill}, \nonumber \\ 
 \Delta \left( {{\bf{R}},{\bf{R}}'} \right) &=& \frac{{4\alpha }}{c}\int\limits_0^\infty  {d\omega } e^{ - i\frac{\omega }{V}\left( {X - X'} \right)} \int\limits_{ - \infty }^{ + \infty } {dq} \int\limits_{ - \infty }^{ + \infty } {dq'} e^{i\frac{\omega }{V}\left( {q - q'} \right)} {\mathop{\rm Im}\nolimits} \left[ {{\cal G}_{\omega xx} \left( {{\bf{r}},{\bf{r}}'} \right)} \right]_{\scriptstyle {\bf{r}} = q{\bf{e}}_x  + Y{\bf{e}}_y  + Z{\bf{e}}_z  \hfill \atop 
  \scriptstyle {\bf{r}}' = q'{\bf{e}}_x  + Y'{\bf{e}}_y  + Z'{\bf{e}}_z  \hfill}.
\end{eqnarray}
By exploiting the rotational invariance of the vacuum longitudinal wavenumber $k_{1z} ( {k_\parallel ^2 } )$ and of the reflection coefficients $R_{\rm IJ} ( {k_\parallel ^2 } )$, from Eqs.(\ref{GGxx}) and (\ref{Gxx}) we get the imaginary part of the $xx$ component of the Green's tensor
\begin{equation}
{\mathop{\rm Im}\nolimits} \left[ {{\cal G}_{\omega xx} \left( {{\bf{r}},{\bf{r}}'} \right)} \right] = \frac{1}{{8\pi ^2 }}\int {d^2 {\bf{k}}_\parallel  e^{i{\bf{k}}_\parallel   \cdot \left( {{\bf{r}}_\parallel   - {\bf{r}}'_\parallel  } \right)} {\mathop{\rm Re}\nolimits} } \left[ {\frac{{e^{ik_{1z} \left| {z - z'} \right|} }}{{k_{1z} }}\left( {1 - \frac{{k_x^2 }}{{k_\omega^2 \varepsilon _1 }}} \right) + \frac{{e^{ - ik_{1z} \left( {z + z'} \right)} }}{{k_{1z} }}\Upsilon } \right]
\end{equation}
where 
\begin{equation}
\Upsilon \left( {\omega ,{\bf{k}}_\parallel  } \right) = \left( {R_{\rm SS} \frac{{k_y^2 }}{{k_\parallel ^2 }} + R_{\rm PP} \frac{{k_{1z}^2 }}{{k_\omega^2 \varepsilon _1 }}\frac{{k_x^2 }}{{k_\parallel ^2 }}} \right) - n\left( {R_{\rm SP} \frac{{k_{1z}^2 }}{{k_\omega^2 \varepsilon _1 }} + R_{\rm PS} } \right)\frac{{k_x k_y }}{{k_\parallel ^2 }}\;
\end{equation}
which inserted into Eqs.(\ref{zeDe0}) yields
\begin{eqnarray} \label{zeDe1}
 \Phi \left( {\bf{R}}_{\rm t} \right) &=& \frac{\alpha }{{2\pi ^2 c}}\int\limits_0^{ + \infty } {d\omega } \int {d^2 {\bf{k}}_\parallel  } \left[ {{\mathop{\rm Im}\nolimits} \int\limits_{ - \infty }^{ + \infty } {dq} \int\limits_{ - \infty }^q {dq'\,} e^{i\left( {k_x  + \frac{\omega }{V}} \right)\left( {q - q'} \right)} } \right]{\mathop{\rm Re}\nolimits} \left[ {\frac{1}{{k_{1z} }}\left( {1 - \frac{{k_x^2 }}{{k_\omega^2 \varepsilon _1 }}} \right) + \frac{{e^{ - i2k_{1z} Z} }}{{k_{1z} }}\Upsilon } \right], \nonumber \\ 
 \Delta \left( {{\bf{R}},{\bf{R}}'} \right) &=& \frac{\alpha }{{2\pi ^2 c}}\int\limits_0^\infty  {d\omega } e^{ - i\frac{\omega }{V}\left( {X - X'} \right)} \int {d^2 {\bf{k}}_\parallel  } e^{ik_y \left( {Y - Y'} \right)} \left[ {\int\limits_{ - \infty }^{ + \infty } {dq} \int\limits_{ - \infty }^{ + \infty } {dq'} e^{i\left( {k_x  + \frac{\omega }{V}} \right)\left( {q - q'} \right)} } \right] \times \nonumber \\
 &\times& {\mathop{\rm Re}\nolimits}\left[ {\frac{{e^{ik_{1z} \left| {Z - Z'} \right|} }}{{k_{1z} }}\left( {1 - \frac{{k_x^2 }}{{k_\omega^2 \varepsilon _1 }}} \right) + \frac{{e^{ - ik_{1z} \left( {Z + Z'} \right)} }}{{k_{1z} }}\Upsilon } \right]. 
\end{eqnarray}
Note that the free-space contribution to $\Phi$ 
is independent on ${\bf R}_{\rm t}$ and hence we hereafter neglect it since it provides an unobservable phase factor in the reduced density matrix. The integrals over $q'$ can be analytically carried out 
\begin{eqnarray}
 {\mathop{\rm Im}\nolimits} \int\limits_{ - \infty }^q {dq'\,} e^{i\left( {k_x  + \frac{\omega }{V}} \right)\left( {q - q'} \right)}  &=& P.V.\frac{1}{{k_x  + \frac{\omega }{V}}}, \nonumber \\ 
 \int\limits_{ - \infty }^{ + \infty } {dq'} e^{i\left( {k_x  + \frac{\omega }{V}} \right)\left( {q - q'} \right)}  &=& 2\pi \delta \left( {k_x  +  \frac{\omega}{V} } \right),
\end{eqnarray}
where $P.V.$ stands for the Cauchy principal value, whereas the remaining integral over $q$ can be replaced by the effective electron path length which can be approximated by $L$, the longitudinal length of the chiral film (see Fig.1 of the main text). By inserting these integrals into Eqs.(\ref{zeDe1}) we get
\begin{eqnarray} \label{zeDe2}
 \Phi \left( {\bf{R}}_{\rm t} \right) &=& \frac{{L\alpha }}{{2\pi ^2 c}}P.V.\int\limits_0^{ + \infty } {d\omega } \;\int {d^2 {\bf{k}}_\parallel  } \frac{1}{{k_x  + \frac{{\omega }}{V }}}{\mathop{\rm Re}\nolimits} \left( {\frac{{e^{ - i2k_{1z} Z} }}{{k_{1z} }}\Upsilon } \right), \nonumber \\ 
 \Delta \left( {{\bf{R}},{\bf{R}}'} \right) &=& \frac{{L\alpha }}{{\pi c}}\int\limits_0^\infty  {d\omega } \int\limits_{ - \infty }^{ + \infty } {dk_y } \;e^{ - i\frac{{\omega }}{V }\left( {X - X'} \right)} e^{ik_y \left( {Y - Y'} \right)} \frac{{e^{\sqrt { k_\omega^2 \left( {\frac{1}{{\beta ^2 }} - \varepsilon _1 } \right) + k_y^2 } \left( {Z + Z'} \right)} }}{\sqrt { k_\omega^2  \left( {\frac{1}{{\beta ^2 }} - \varepsilon _1 } \right) + k_y^2 }}{\mathop{\rm Im}\nolimits} \left( \Upsilon  \right)_{k_x  =  - \frac{{\omega }}{V }}.
\end{eqnarray}
Note that the free-space part of the Green's tensor does not affect $\Delta \left( {{\bf{R}},{\bf{R}}'} \right)$ since its contribution in the curly bracket in the second of Eqs.(\ref{zeDe1}) is purely imaginary because $k_{1z}= i\sqrt { k_\omega^2  \left( {\frac{1}{{\beta ^2 }} - \varepsilon _1 } \right) + k_y^2 }$ for $k_x = - \omega /V$. In addition, due to the planar geometry of the chiral film, the phase factor $\Phi$ is a function only of $Z$ whereas $\Delta \left( {{\bf{R}},{\bf{R}}'} \right)$ only depends on $X - X'$, $Y - Y'$ and $Z + Z'$. Equations (\ref{zeDe2}) enable the evaluation of the  reduced density matrix of Eq.(\ref{rhoe}).

\subsection{Mirror symmetry breaking effects produced by mirror optical activity}
The Green's tensor component $xx$ is not symmetric under mirror reflections owing to MOA, as discussed above, and this entails that the aloof electron-chiral film interaction has broken mirror symmetry. In order to discuss this crucial point, from the second Eqs.(\ref{zeDe2}) it is convenient noting that the decomposition
\begin{equation} \label{DeSpA}
\Delta \left( {{\bf{R}},{\bf{R}}'} \right) = \Delta _{\rm S} \left( {X - X',Y - Y',Z + Z'} \right) + i \Delta _{\rm A} \left( {X - X',Y - Y',Z + Z'} \right)
\end{equation}
holds, in which
\begin{eqnarray} \label{DeSA}
 \Delta _{\rm S}  ( {\tilde X,\tilde Y,\tilde Z}  ) &=& \frac{{L\alpha }}{{\pi c}}\int\limits_0^\infty  {d\omega } \;e^{ - i\frac{\omega}{V }\tilde X} \int\limits_{ - \infty }^{ + \infty } {dk_y } \;\cos  ( {k_y \tilde Y}  )\frac{{e^{\sqrt { k_\omega^2  \left( {\frac{1}{{\beta ^2 }} - \varepsilon _1 } \right) + k_y^2 } \tilde Z} }}{\sqrt { k_\omega^2  \left( {\frac{1}{{\beta ^2 }} - \varepsilon _1 } \right) + k_y^2 }}{\mathop{\rm Im}\nolimits} \left( {R_{\rm SS} \frac{{k_y^2 }}{{k_\parallel ^2 }} + R_{\rm PP} \frac{{k_{1z}^2 }}{{k_0^2 \varepsilon _1 }}\frac{{k_x^2 }}{{k_\parallel ^2 }}} \right)_{k_x  =  - \frac{{\omega}}{V }}, \nonumber  \\ 
 \Delta _{\rm A}  ( {\tilde X,\tilde Y,\tilde Z}  ) &=& \frac{{L\alpha }}{{\pi c}}\int\limits_0^\infty  {d\omega } \;e^{ - i\frac{\omega}{V }\tilde X} \int\limits_{ - \infty }^{ + \infty } {dk_y } \sin  ( {k_y \tilde Y}  )\frac{{e^{\sqrt { k_\omega^2  \left( {\frac{1}{{\beta ^2 }} - \varepsilon _1 } \right) + k_y^2 } \tilde Z} }}{\sqrt { k_\omega^2  \left( {\frac{1}{{\beta ^2 }} - \varepsilon _1 } \right) + k_y^2 }}{\mathop{\rm Im}\nolimits} \left[ { - n\left( {R_{\rm SP} \frac{{k_{1z}^2 }}{{k_0^2 \varepsilon _1 }} + R_{\rm PS} } \right)\frac{{k_x k_y }}{{k_\parallel ^2 }}} \right]_{k_x  =  - \frac{{\omega }}{V }} , \nonumber \\
\end{eqnarray}
are symmetric and antisymmetric under the reflection through the $xz$ plane, respectively, since 
\begin{eqnarray}
 \Delta _{\rm S}  ( {\tilde X, - \tilde Y,\tilde Z}  ) &=& \Delta _{\rm S}  ( {\tilde X,\tilde Y,\tilde Z}  ), \nonumber \\ 
 \Delta _{\rm A}  ( {\tilde X, - \tilde Y,\tilde Z}  ) &=&  - \Delta _{\rm A}  ( {\tilde X,\tilde Y,\tilde Z}  ).
\end{eqnarray}
Note that $\Delta _{\rm A}$ is entirely due to MOA since it only contains the mixing reflection coefficients $R_{\rm SP}$ and $R_{\rm PS}$ and accordingly its sign is switched by chirality reversal $\kappa \rightarrow -\kappa$ since it is proportional to the chiral refractive index $n$. Equation (\ref{DeSpA}) shows that the mirror symmetry of $\Delta$ is broken by MOA and it enables the decoherence factor of Eq.(\ref{gammmm}) to be written as
\begin{equation} \label{gammFILM}
\gamma \left( {{\bf{R}},{\bf{R}}'} \right) = e^{i\left[ {\Phi \left( Z \right) - \Phi \left( {Z'} \right)} \right]} e^{ \Delta _{\rm S} \left( {X - X',Y - Y',Z + Z'} \right) - \frac{1}{2}\left[ {\Delta _{\rm S} \left( {0,0,2Z} \right) + \Delta _{\rm S} \left( {0,0,2Z'} \right)} \right]  } e^{i \Delta _{\rm A} \left( {X - X',Y - Y',Z + Z'} \right)}  
\end{equation}
which, in agreement with the general case of Eq.(\ref{ASMgam}), evidently has broken mirror symmetry here due the last exponential factor containing $\Delta _{\rm A}$. In order to quantify its absolute mirror asymmetry we consider the dimensionless quantity
\begin{equation} \label{asym1}
{\rm Asym} \left[ {\gamma \left( {{\bf{R}},{\bf{R}}'} \right)} \right] = 2\frac{{\gamma \left( {{\bf{R}},{\bf{R}}'} \right) - \gamma \left( {{\cal R}{\bf{R}},{\cal R}{\bf{R}}'} \right)}}{{\gamma \left( {{\bf{R}},{\bf{R}}'} \right) + \gamma \left( {{\cal R}{\bf{R}},{\cal R}{\bf{R}}'} \right)}}
\end{equation}
which compares $\gamma\left( {{\bf{R}},{\bf{R}}'} \right)$ and its mirror image $\gamma \left( {{\cal R}{\bf{R}},{\cal R}{\bf{R}}'} \right)$ through the ratio between their difference and their arithmetic average. Inserting Eq.(\ref{gammFILM}) into Eq.(\ref{asym1}) we obtain
\begin{equation} \label{asym2}
{\rm Asym} \left[ {\gamma \left( {{\bf{R}},{\bf{R}}'} \right)} \right] = 2i \tan \left[  {\Delta _{\rm A} \left( {X-X',Y-Y',Z+Z'} \right)} \right]
\end{equation}
clearly elucidating that MOA breaks the mirror symmetry of the electron decoherence. Since $\Delta_{\rm A}$ is proportional to $L$, the absolute asymmetry of the decoherence factor can be enhanced in the presence of long effective interaction lengths. For an initial mirror symmetric electron wave function (see Eq.(\ref{phiiMIS})), Eq.(\ref{asym2}) also provides the absolute mirror asymmetry of the electron reduced density matrix (see Eq.(\ref{rhoe})).

As discussed in Sec.2.5, the electron transverse momentum distribution lacks mirror symmetry along the lateral direction. In order to relate such general effect to MOA in the configuration we are examing, we here focus on the distribution of the lateral momentum component $P_y$ which is given by  $\frac{{d\mathscr{P}}}{{dP_y }} = \int {dP_z } \frac{{d\mathscr{P}}}{{d^2 {\bf{P}}_{\rm t} }}$. From the first of Eqs.(\ref{dP3}), we  straightforwardly get
\begin{equation}
\frac{{d \mathscr{P}}}{{dP_y }} = \frac{1}{{2\pi \hbar }}\int {dY} \int {dY'} \int {dZ} \, e^{ - \frac{i}{\hbar }P_y \left( {Y - Y'} \right)}  \phi _i \left( {Y,Z} \right)\phi _i^* \left( {Y',Z} \right)\left[ {\gamma \left( {{\bf{R}}_{\rm t} ,{\bf{R}}'_{\rm t} } \right)} \right]_{Z' = Z}
\end{equation}
which, after using Eq.(\ref{gammFILM}) and performing the change of integration variables $Y' = Y - \tilde Y$, $Z = \tilde Z/2$, becomes
\begin{equation} \label{dPyD}
\frac{{d \mathscr{P}}}{{dP_y }} = \frac{1}{{4\pi \hbar }}\int {d\tilde Y} \int {d\tilde Z} \, e^{ - \frac{i}{\hbar }P_y \tilde Y}  \tilde \phi _i  ( {\tilde Y,\tilde Z}  ) e^{\Delta _{\rm S} \left( {0,\tilde Y,\tilde Z} \right) - \Delta _{\rm S} \left( {0,0,\tilde Z} \right)} e^{i\Delta _{\rm A} \left( {0,\tilde Y,\tilde Z} \right)} 
\end{equation}
where 
\begin{equation}
\tilde \phi _i ( {\tilde Y,\tilde Z} ) = \int {dY} \,\phi _i  ( {Y,{\tilde Z}/2}  )\phi _i^*  ( {Y - \tilde Y,{\tilde Z}/2}  )
\end{equation}
is the autoconvolution of the initial electron wavefunction. For the initial mirror symmetric electron wave function $\phi _i \left( { - Y,Z} \right) = \phi _i \left( {Y,Z} \right)$, the autoconvolution $\tilde \phi _i  ( {  \tilde Y,\tilde Z}  ) $ is even in $\tilde Y$. Now the real quantities $\Delta _{\rm S}  ( {0,\tilde Y,\tilde Z}  ) - \Delta _{\rm S}  ( {0,0,\tilde Z}  )$ and $\Delta _{\rm A} ( {0,\tilde Y,\tilde Z} )$ are  even and odd in $\tilde Y$, respectively, and hence we conclude that $\frac{{d \mathscr{P}}}{{dP_y }}$ of Eq.(\ref{dPyD}) can not be either even or odd in $P_y$ since it is the Fourier transform of a function whose parity is indefinite. To explicitly highlight its asymmetry we write Eq.(\ref{dPyD}) as
\begin{eqnarray}
\frac{{d \mathscr{P}}}{{dP_y }} &=& \frac{1}{{4\pi \hbar }}\int {d\tilde Y} \int {d\tilde Z} \;\tilde \phi _i  ( {\tilde Y,\tilde Z}  ) e^{\Delta _{\rm S} \left( {0,\tilde Y,\tilde Z} \right) - \Delta _{\rm S} \left( {0,0,\tilde Z} \right)} \times \nonumber \\
&\times& \left\{ \underbrace{ \cos \left( {\frac{1 }{\hbar } P_y\tilde Y} \right)\cos \left[ {\Delta _{\rm A}  ( {0,\tilde Y,\tilde Z}  )} \right] 
}_{{\rm even\,in}\,P_y } + \underbrace{\sin \left( {\frac{1}{\hbar }P_y \tilde Y} \right)\sin \left[ {\Delta _{\rm A}  ( {0,\tilde Y,\tilde Z}  )} \right]}_{{\rm odd\,in} \,P_y } \right\}
\end{eqnarray}
where the even and odd contributions to the transverse momentum distribution are separeted. The odd part is entirely due to $\Delta_{\rm A}$ so that the mirror symmetry of the electron lateral momentum distribution is broken by MOA.

As discussed in Sec.S2.5, the asymmetry of the transverse momentum distribution has the remarkable consequence that the electron gains a net lateral momentum upon scattering. To explicitly discuss such phenomenon for the aloof electron we are considering, we evaluate the mean value and variance of $P_y$. After substituting Eq.(\ref{DeSpA}) into Eq.(\ref{THETA}), and using the fact that the impinging electron with mirror symmetric wavefuction has no transverse momentum, $\int dY \int dZ \phi _i^* \left( {\frac{\hbar }{i}\frac{\partial }{{\partial Y}}} \right)\phi _i  = 0$, the firsts of Eqs.(\ref{meanvalues}) and Eqs.(\ref{variances}) with $j=y$, after some algebra, yield
\begin{eqnarray} \label{Pmeanvariance}
 \left\langle {P_y } \right\rangle  &=& \int dY \int dZ \;\left| {\phi _i } \right|^2 A, \nonumber \\ 
 \left\langle {P_y^2 } \right\rangle  - \left\langle {P_y } \right\rangle ^2  &=& \int dY \int dZ \phi _i^* \left( {\frac{\hbar }{i}\frac{\partial }{{\partial Y}}} \right)^2 \phi _i  + \int dY \int dZ \left| {\phi _i } \right|^2 S + \left\{ {\int dY \int dZ \left| {\phi _i } \right|^2 A^2  - \left[ {\int dY \int dZ \left| {\phi _i } \right|^2 A} \right]^2 } \right\}, \nonumber \\
\end{eqnarray}
where 
\begin{eqnarray}
 A \left( Z \right) &=& \left[ {\hbar \frac{{\partial \Delta _{\rm A}  ( {0,\tilde Y,2Z}  )}}{{\partial \tilde Y}}} \right]_{\tilde Y = 0}, \nonumber  \\ 
 S \left( Z \right) &=& \left[ { - \hbar ^2 \frac{{\partial ^2 \Delta _{\rm S}   ( {0,\tilde Y,2Z}  )}}{{\partial \tilde Y^2 }}} \right]_{\tilde Y = 0}.
\end{eqnarray}
The first of Eqs.(\ref{Pmeanvariance}) makes explicit that the mean value of $P_y$ is fully due to MOA, as expected and it is one of the main results of this work. The second of Eqs.(\ref{Pmeanvariance}) revelas that the variance of $P_y$ has two contributions that add up to the initial variance $\int dY \int dZ \phi _i^* \left( {\frac{\hbar }{i}\frac{\partial }{{\partial Y}}} \right)^2 \phi _i$. The one containing $\Delta_{\rm S}$ (through $S$) is not vanishing even in the achiral limit $\Delta_{\rm A}=0$ and it accounts for the natural momentum distribution reshaping produced by the interaction of the electron with the medium. On thet other hand, the contribution containing $\Delta_{\rm A}$ (through $A$) is entirely produced by MOA and it is expected to be very small as compared to the previous one.

\subsection{Electron energy mean value and variance}
As discussed in section S2.5, the interaction of the electron with the chiral medium results from the combined effect of an infinite number of processes in each of which the energy lost by the electron is transferred to a finite number of polaritonic excitations. As a consequence, the energy of the electron after the interaction is not definite and it has the probability distribution discussed in section S2.4 which, for the aloof electron we are considering, can be obtained by substituing the decoherence factor of Eq.(\ref{gammFILM}) into the second of Eqs.(\ref{dP3}), thus getting
\begin{equation}
\frac{{d \mathscr{P}}}{{dE}} = \frac{1}{{2\pi \hbar V}}\int {dY} \int {dZ} \int {dQ} e^{ - i\frac{{E - E_i }}{{\hbar V}}Q} \left| {\phi _i \left( {Y,Z} \right)} \right|^2 e^{\Delta _{\rm S} \left( {Q,0,2Z} \right) - \Delta _{\rm S} \left( {0,0,2Z} \right)}.
\end{equation}
This expression does not contain $\Delta_{\rm A}$ and hence we conclude that the electron-medium energy exchance is not affected by MOA, as expected since the electron energy is invariant under mirror reflection. In analogy with the above discussed transverse momentum, after substituting Eq.(\ref{DeSpA}) into Eq.(\ref{THETA}), the seconds of Eqs.(\ref{meanvalues}) and Eqs.(\ref{variances}), after some algebra, yield the energy mean value and variance
\begin{eqnarray} \label{Emeanvariance}
\left\langle E \right\rangle  &=& E_i  + \int dY \int dZ \left| {\phi _i } \right|^2 \sigma ^{\left( 1 \right)}, \nonumber  \\ 
 \left\langle {E^2 } \right\rangle  - \left\langle E \right\rangle ^2  &=& \int dY \int dZ \left| {\phi _i } \right|^2 \sigma ^{\left( 2 \right)}  + \left\{ {\int {dY dZ} \left| {\phi _i } \right|^2 \left( {\sigma ^{\left( 1 \right)} } \right)^2  - \left[ {\int dY \int dZ \left| {\phi _i } \right|^2 \sigma ^{\left( 1 \right)} } \right]^2 } \right\} 
\end{eqnarray}
where
\begin{equation}
\sigma ^{\left( n \right)} \left( Z \right) = \left[ {\left( {V\frac{\hbar }{i}\frac{\partial }{{\partial \tilde X}}} \right)^n \Delta _{\rm S} ( {\tilde X,0,2Z} )} \right]_{\tilde X = 0}.
\end{equation}
Using the first of Eqs.(\ref{DeSA}) it is simple proving that $\sigma ^{\left( 1 \right)} \left( Z \right) < 0$ so that the first of Eqs.(\ref{Emeanvariance}) shows that the energy mean value is smaller than the initial energy $E_i$, as expected since part of the initial electron energy is delivered to the field.